\documentclass[preprint,amsmath,amssymb,superscriptaddress,nofootinbib,showpacs,preprintnumbers, keywords]{revtex4}

\usepackage{bm}
\usepackage{booktabs}
\usepackage{mathrsfs}
\usepackage[pdftex]{graphicx,color}
\usepackage{epstopdf}
\usepackage[bookmarks=true,bookmarksnumbered=true,bookmarkstype=toc]{hyperref} 
\usepackage{units}

\usepackage{mathtools}

\usepackage[utf8]{inputenc}
\usepackage{color}

\begin{document}
\title{
Big-bang nucleosynthesis and Leptogenesis in CMSSM
}

\author{Munehiro Kubo
}
\email{kubo@krishna.th.phy.saitama-u.ac.jp}
\affiliation{%
Department of Physics, Saitama University,
\\
Shimo-Okubo 255, 
338-8570 Saitama Sakura-ku, 
Japan
}

\author{Joe Sato}
\email{joe@phy.saitama-u.ac.jp}
\affiliation{%
Department of Physics, Saitama University,
\\
Shimo-Okubo 255, 
338-8570 Saitama Sakura-ku, 
Japan
}

\author{Takashi Shimomura}
\email{shimomura@cc.miyazaki-u.ac.jp}
\affiliation{%
Faculty of Education, 
University of Miyazaki,
\\
Gakuen-Kibanadai-Nishi 1-1,
889-2192 Miyazaki,
Japan
}

\author{Yasutaka Takanishi}
\email{yasutaka@krishna.th.phy.saitama-u.ac.jp}
\affiliation{%
Department of Physics, Saitama University,
\\
Shimo-Okubo 255, 
338-8570 Saitama Sakura-ku, 
Japan
}

\author{Masato Yamanaka}
\email{masato.yamanaka@cc.kyoto-su.ac.jp}
\affiliation{%
Maskawa Institute, Kyoto Sangyo University, Kyoto 603-8555, Japan
}

\begin{abstract}
  We have investigated the constrained minimal supersymmetric standard
  model with three right-handed Majorana neutrinos whether there still
  is a parameter region which is consistent with all existing
  experimental data/limits such as Leptogenesis and the dark matter
  abundance and we also can solve the Lithium problem. 
  Using Casas-Ibarra parameterization, we have found that a very
  narrow parameter space of the complex orthogonal matrix elements
  where the lightest slepton can have a long lifetime, that is
  necessary for solving the Lithium problem. Further, under this
  condition, there is a parameter region that can give an explanation
  for the experimental observations. We have studied three cases of
  the right-handed neutrino mass ratio \mbox{\em (i)}
  $M_{2}=2 \times M_{1}$, \mbox{\em (ii)}
  $M_{2}=4 \times M_{1}$, \mbox{\em (iii)}
  $M_{2}=10 \times M_{1}$ while $M_{3}=40 \times M_{1}$ is fixed.
   We have obtained the
  mass range of the lightest right-handed neutrino mass that lies
  between $10^9$~GeV and $10^{11}$~GeV. The important result is that
  its upper limit is derived by solving the Lithium problem and the
  lower limit comes from Leptogenesis.
  Calculated low-energy observables of these parameter sets such as
  BR($\mu \to e \gamma$) is not yet restricted by experiments and will
  be verified in the near future.
\end{abstract}

\date{\today}



\preprint{\bf STUPP-18-233, UME-PP-009, MISC-2018-01}

\maketitle

\section{Introduction} \label{sec:intro}

The standard models (SMs) of particle physics and cosmology have been
successful to understand most of experimental and observational
results obtained so far. Nonetheless, there are several phenomena
which cannot be explained by these models.  Among such phenomena, the
mass and mixing of neutrinos, the Baryon asymmetry of the universe
(BAU), the existence of the dark matter (DM), so-called Lithium (Li)
problems are compelling evidences that require new physics laws for
explanations.  If all of these phenomena are addressed in particle
physics, the new physics laws should be incorporated in a unified
picture beyond the SM of particle physics.

Neutrino oscillation experiments (see Ref.~\cite{deSalas:2017kay} for
recent review and global fit analysis) and cosmological observations
\cite{Ade:2015xua, Aghanim:2016yuo} revealed that the masses of
neutrinos are much lighter than those of other known SM particles. To
generate such tiny masses many mechanisms have been proposed, among
which most well-studied and the simplest mechanism is (type-I) seesaw
mechanism \cite{Minkowski:1977sc, Yanagida:1979as, GellMann:1980vs,
  Glashow:1979nm, Mohapatra:1979ia}.  In this mechanism, the heavy
Majorana right-handed (RH) neutrinos are introduced and thus the
Yukawa interactions of left-handed (LH) and RH neutrinos can be formed
with the Higgs scalar, that gives rise to the flavor mixings in the
neutrino sector.  After integrating out the RH neutrinos, the LH
neutrino masses become very light due to the suppression factor which
is proportional to the inverse of the Majorana mass scale. Thus when we
make use of the seesaw mechanics we can successfully generate the
phenomenologically required masses and mixings of low-energy LH
neutrinos.

Furthermore, the seesaw mechanism has another virtue, generating the
baryon asymmetry \cite{Ade:2015xua} through Leptogenesis
\cite{Fukugita:1986hr}.  At the early stage of the universe, the RH
Majorana neutrinos are produced in the thermal bath. As temperature
decreases to their mass scale these neutrinos go out-of-thermal
equilibrium, and at that time they decay into lepton with Higgs
or anti-lepton with anti-Higgs.  If CP symmetry is violated in the
neutrino Yukawa coupling, the decay rates into lepton and anti-lepton are obviously different. 
That means that the lepton number asymmetry is generated through the decays of
the heavy Majorana RH neutrinos, and then the lepton number asymmetry is converted
to the baryon asymmetry by sphaleron process \cite{Kuzmin:1985mm,
  Harvey:1990qw}: the seesaw mechanism explains two phenomena
simultaneously. (see \mbox{\it e.g.} Refs.~\cite{Buchmuller:2002rq,Ellis:2002xg, Bando:2004hi,Chang:2004wy,Petcov:2005jh,
Guo:2006qa,Pascoli:2006ci})

The existence of DM is also problem \cite{Spergel:2006hy}.  
The dark matter must be a massive and stable or very long-lived particle
compared with the age of the universe and do not carry electric
neither color charges.  Neutrino is only possible candidate for the DM
within the SM, however, this possibility has been already ruled out
because neutrino masses are too light. Thus, one should extend the SM
so that the DM is incorporated.
Supersymmetry (SUSY) with $R$ parity is one of the attractive
extensions in this regard, where the lightest SUSY particle (LSP)
becomes absolutely stable. In many SUSY models, the LSP is the
lightest neutralino that is a linear combination of neutral components
of gauginos and higgsinos that are SUSY partners of electroweak gauge
bosons and the Higgses, respectively.  Therefore, the lightest
neutralino LSP is a good candidate for the DM, and in fact the
abundance of the neutralino LSP can be consistent with observational
one of the DM \cite{Spergel:2006hy} in specific parameter regions.
In particular, the so-called coannihilation region is very
interesting, in which the neutralino DM and the lighter stau, SUSY
partner of tau lepton, as the next-LSP (NLSP) are degenerate in mass
\cite{Griest:1990kh}.  When the mass difference of the neutralino LSP
and the stau NLSP is smaller than $\mathcal{O}(100)$ MeV, the stau
NLSP becomes long-lived so that it can survive during the Big-Bang
nucleosynthesis (BBN) proceeds \cite{Profumo:2004qt,Gladyshev:2005mn, Jittoh:2005pq}.  Thus, the
existence of the stau NLSP affects the primordial abundance of light
elements.  One can expect to find evidences of the stau NLSP in primordial
abundance of light elements.

It has been reported that there are disagreements on the primordial
abundances of $^7$Li and $^6$Li between the standard BBN prediction
and observations. The prediction of the $^7$Li abundance is about $3$
times larger than the observational one
$(1.6 \pm 0.3) \times 10^{-10}$ \cite{Cyburt:2008kw, Sbordone:2010zi,
  Coc:2011az}.  This discrepancy hardly seems to be solved in the standard BBN 
  with the measurement errors. 
 This is called the $^7$Li problem. The $^6$Li
abundance is also disagreed with the observations.  The predicted
abundance is about $10^3$ smaller than the observational abundance
$^6$Li/$^7$Li~$\simeq 5 \times 10^{-2}$ \cite{Asplund:2005yt}.
Although this disagreement is less robust because of uncertainties of
theoretical prediction, it is called the $^6$Li problem.

Since the disagreements cannot attribute to nuclear physics in the BBN \cite{Kawabata:2017zpa}, 
one needs to modify the standard BBN reactions.  In Ref.~\cite{Jittoh:2007fr}, the
authors have shown in the minimal SUSY standard model (MSSM) that negatively
charged stau can form bound states with light nuclei, and immediately
destroy the nuclei through the internal conversion processes during the
BBN. Further, a detail analysis \cite{Jittoh:2008eq} has showed that in
the coannihilation region where the lightest neutralino LSP is the DM
and the stau NLSP has lifetime of $\mathcal{O}(10^3)$ sec., Li and
Beryllium (Be) nuclei are effectively destroyed.  The primordial density
of $^7$Li is reduced, while such a stau can promote to produce $^6$Li
density~\cite{Pospelov:2006sc}. It turns out that both densities become
the observational values.  This is a solution of the dark matter and the
Li problems in the MSSM scenario.  It should be noted that the SUSY
spectrum is highly predictive in this parameter region.  In
Ref.~\cite{Konishi:2013gda}, the authors also showed whole SUSY spectrum
in which the lightest neutralino mass is between 350 GeV and 420
GeV in the constrained MSSM (CMSSM).  This result is consistent with
non-observation of SUSY particle at the LHC experiment so far. 
However it is in the reach of the LHC Run-II.

In this article, we consider the CMSSM with the type I seesaw
mechanism as a unified picture which successfully explains all
phenomena as we have mentioned above. We aim to examine this model
through searches of the long-lived charged particles at the LHC and
lepton flavor violation  (see \mbox{\it e.g.} Refs.~\cite{Borzumati:1986qx,Hisano:1995cp,Casas:2001sr,Ellis:2001xt,Ellis:2002fe,Lavignac:2001vp,Kageyama:2001tn,%
Deppisch:2002vz,Blazek:2002wq,Petcov:2003zb,Dutta:2003ps,Illana:2003pj,Bando:2004hi,Babu:2005yr,Petcov:2005yh,Calibbi:2017uvl})  
at MEG-II, Mu3e and Belle-II experiments. This paper is organized as follows. In section II,
we review the CMSSM with the heavy RH Majorana neutrinos.  In section
III, we show cosmological constraints such as dark matter, BBN and
BAU, which we require to the model in our analysis. Then, we present
the parameter sets of the CMSSM and RH Yukawa coupling which satisfies
all requirements in section IV. Predictions on lepton flavor
violating decays are shown in Section V.  The last section is devoted
to summary and discussion.

\section{Model and Notation}
\label{sec:notation}

We consider the MSSM with RH Majorana neutrinos (MSSMRN).  The superpotential for the lepton sector
is given by
\begin{align}
\mathscr{W}_l &=\widehat{E}^{c}_\alpha \left(Y_{\rm E} \right)_{\alpha\beta} \widehat{L}_\beta \cdot
  \widehat{H}_d + 
\lambda_{\beta i} \widehat{L}_\beta \cdot \widehat{H}_u \widehat{N}^c_i 
- \frac{1}{2} \left( M_{\rm N} \right)_{ij}\widehat{N}^c_i \widehat{N}^c_j \hspace{2mm}.
\label{eq_sptl}
\end{align}
%
Here $\widehat{L}_\alpha$ and $\widehat{E}^c_\alpha$
($\alpha=e,\mu,\tau; i, j = 1,2,3$), are the chiral supermultiplets
respectively of the $SU(2)_L$ doublet lepton and of the $SU(2)_L$
singlet charged lepton in the flavor basis which is given as the mass
eigenstate of the charged lepton, that is, the eigenstate of $Y_E$ and
hence implicitly $(Y_E)_{\alpha\beta}=y_\alpha\delta_{\alpha\beta}$ is
assumed. Similarly $\widehat{N}_{i}^c\, (i=1,2,3)$ is that of
RH neutrino and indices denote the mass eigenstate, that is,
the eigenstate of $M_N$ and implicitly ${M_N}_{ij}=M_i\delta_{ij}$ is
assumed and  the superscript $C$ denotes the
charge conjugation. 
$\widehat{H}_u$ and $\widehat{H}_d$ are the supermultiplets of
the two Higgs doublet fields $H_u$ and $H_d$.

Below the lightest RH seesaw mass scale, the singlet supermultiplets
$\widehat{N}^c_i$ containing the RH neutrino fields are integrated
out, the Majorana mass term for the LH neutrinos in the
flavor basis is obtained
\begin{eqnarray}
\mathscr{L}_{m}^{\nu} &=& 
- \frac{1}{2}~\nu_{L\alpha} \left(m_{\nu}\right)_{\alpha\beta} \nu_{L\beta} +
h.c.~,\\
\left( m_{\nu} \right)_{\alpha \beta} &=&  v_{u}^2~\left(
\lambda_{\nu}\right)_{\alpha i} M_{i}^{-1} \left( \lambda_{\nu}
\right)_{i\beta} \hspace{2mm},
\label{mnuKN}
\end{eqnarray}
%
%
where $M_i = (M_1, M_2, M_3)$ and $v_u$ is vacuum expectation value
(VEV) of up-type Higgs field $H_{u}$, $v_u = v \sin\beta$ with
$v=174$~GeV\@. The matrix $\left( m_{\nu} \right)_{\alpha \beta}$ can be
diagonalized by a single unitary matrix -- Maki-Nakagawa-Sakata-matrix --
$U_{\mathrm{MNS}}$ as
\begin{equation}
\left( m_{\nu} \right) = U^{*}_{\mathrm{MNS}} ~ \mathrm{D}_{m_{\nu}} ~ {U^{\dagger}}_{\mathrm{MNS}} \hspace{2mm},
\end{equation}
where $\mathrm{D}_{m_{\nu}}=\mathrm{diag}(m_{\nu_{1}},m_{\nu_{2}},m_{\nu_{3}})$.

The solar, atmospheric and reactor neutrino experiments have shown at 3 $\sigma$ level 
that~\cite{Patrignani:2016xqp}
\begin{eqnarray}
\label{eq:experimvalues}
&& \Delta m_{12}^2 =(6.93 - 7.96)\times10^{-5}~\mathrm{(eV^2)}\,, \quad 
\Delta m_{23}^2 =(2.42 - 2.66)\times10^{-3}~\mathrm{(eV^2)}\,,\nonumber\\
&&\!\sin^2\theta_{12} = (0.250 - 0.354)  , \quad 
\!\sin^2\theta_{\rm 23} = (0.381 - 0.615) , \quad 
\!\!\sin^2\theta_{13} = (0.0190 - 0.0240) \,.
\end{eqnarray}
Note that in this article we will assume that the mass spectrum of
light neutrinos is hierarchical ($m_{\nu_1} \ll m_{\nu_2}\ll m_{\nu_3}$) and thus
$m_{\nu_3}\simeq\sqrt{\Delta m^2_{\rm atm}}$ and $m_{\nu_2}\simeq\sqrt{\Delta
  m^2_\odot}$ and also that all mixing angles lie in the interval
$0<\theta_{12},\theta_{23},\theta_{13}<\pi/2$. Furthermore, the
lightest LH neutrino mass is fixed , for our main result, to be
\begin{equation}
m_{\nu_{1}}=0.001~\mathrm{(eV)} \hspace{2mm},
\end{equation}
as we will see that we have no solution of degenerate case.

We will use the standard parametrization of the
MNS matrix
\begin{equation}
U_{\mathrm{MNS}}= \widehat{U} ~ \mathrm{diag} \left(1, e^{i\alpha}, e^{i\beta} \right) \hspace{2mm},
\end{equation}
with
\begin{equation}
\label{eq:MNS}
\widehat{U} = \left(
    \begin{array}{ccc}
c_{13}c_{12} & c_{13}s_{12} & s_{13}e^{-i\delta} \\
-c_{23}s_{12}-s_{23}s_{13}c_{12}e^{i\delta} &
c_{23}c_{12}-s_{23}s_{13}s_{12}e^{i\delta} & s_{23}c_{13} \\
s_{23}s_{12}-c_{23}s_{13}c_{12}e^{i\delta} &
-s_{23}c_{12}-c_{23}s_{13}s_{12}e^{i\delta} & c_{23}c_{13}
\end{array}
  \right) \hspace{2mm},
\end{equation}
\noindent where $c_{ij} = \cos\theta_{ij}$, $s_{ij} =
\sin\theta_{ij}$, and $\delta$ is the Dirac CP-violating phase and $\alpha$
and $\beta$ are two Majorana CP-violation phases. The input values of
the angles and three CP-violation phases at GUT scale are set
respectively by
\begin{eqnarray}
\label{eq:expmixings}
&&s_{23} = \sqrt{0.441} \,,\, s_{13} = \sqrt{0.02166} \,,\, s_{12} = \sqrt{0.306} \,, \nonumber\\ 
&&\alpha = 0 \,, \, \beta = 0 \,,  \delta = 261^{\circ} \hspace{2mm}.
\end{eqnarray}

In addition, we parameterize the matrix of neutrino Yukawa couplings a la Casas-Ibarra~\cite{Casas:2001sr}
\begin{equation}
\label{eq:eq9}
\lambda_{\nu}=\frac{1}{v_{u}} ~ U_{\mathrm{MNS}}^{*}\sqrt{\!\hspace{1mm}\mathrm{D}_{m_{\nu}}}~ R ~ \sqrt{M} \hspace{2mm}, 
\end{equation}
where %
\begin{equation}
  R=\left(
    \begin{array}{ccc}
      \widetilde{c}_{13}\widetilde{c}_{12} & \widetilde{c}_{13}\widetilde{s}_{12} & \widetilde{s}_{13} \\
      -\widetilde{c}_{23}\widetilde{s}_{12} & \widetilde{c}_{23}\widetilde{c}_{12}-\widetilde{s}_{23}\widetilde{s}_{13}\widetilde{s}_{12} & \widetilde{s}_{23}\widetilde{c}_{13} \\
      \widetilde{s}_{23}\widetilde{s}_{12}-\widetilde{c}_{23}\widetilde{s}_{13}\widetilde{c}_{12} & -\widetilde{s}_{23}\widetilde{c}_{12}-\widetilde{c}_{23}\widetilde{s}_{13}\widetilde{s}_{12} & \widetilde{c}_{23}\widetilde{c}_{13}
    \end{array}
  \right) \hspace{2mm}.
\end{equation}
We adopt that $R$ is a complex orthogonal matrix, $R^{\bf T}R = 1$, so
that $\widetilde{c}_{i j}=\cos{z_{i j}}$ and
$\widetilde{s}_{i j}=\sin{z_{i j}}$ with
$z_{i j} = x_{i j} + \sqrt{-1}~y_{i j}$ because we will calculate
CP-violating process such as Leptogensis.

\section{Cosmological constraint} 
\label{sec:cosmo}

For our analysis we take into account three types of cosmological
observables; \mbox{\it (i)} dark matter abundance \mbox{\it (ii)}
light element abundances \mbox{\it (iii)} baryon asymmetry of the
universe.
We show our strategy to find favored parameter space from a standpoint
of these observables in the CMSSM with seesaw mechanism.

\subsection{Number densities of dark matter and of long-lived slepton} 
\label{subsec:density}

We consider the neutralino-slepton coannihilation scenario in the framework of CMSSM 
wherein the LSP is the Bino-like neutralino
$\widetilde{\chi}_{1}^{0}$ and the NLSP is the lightest
slepton $\widetilde{\ell}_{1}$ that almost consists of RH stau 
including tiny flavor mixing, 
\begin{equation}
\begin{split}
	\widetilde{\ell}_{1} 
	&= 
	\sum_{f=e, \mu, \tau} C_{f} \widetilde{f}, 
\label{Eq:slepton}
\end{split}     
\end{equation}
where $C_{e}^{2}+C_{\mu}^{2}+C_{\tau}^{2}=1$, and 
each interaction state is 
\begin{equation}
\begin{split}
	\widetilde{f} 
	&= 
	\cos\theta_{f} \widetilde{f}_{L} + \sin\theta_{f} \widetilde{f}_{R}\,.
\label{Eq:f}
\end{split}     
\end{equation}
The flavor mixing $C_{f}$ and left-right mixing angle 
$\theta_{f}$ are determined by solving RG equations with 
neutrino Yukawa. 
In our scenario 
$C_\tau \sim 1 \gg 
C_e, C_\mu$ and $\sin\theta_\tau\sim 1$.

The standard calculation for relic density of the 
$\widetilde{\chi}_{1}^{0}$ leads to an over-abundant dark 
matter density. 
A tight mass degeneracy between $\widetilde{\ell}_{1}$ and 
$\widetilde{\chi}_{1}^{0}$ assists to maintain the chemical 
equilibrium of SUSY particles with SM sector, and can reduce the 
relic density below the Planck bound~\cite{Aghanim:2016yuo}. 
This is called coannihilation mechanism~\cite{Griest:1990kh}.

In a unique parameter space for the neutralino-slepton 
coannihilation to work well, we focus on the space 
where the mass difference between $\widetilde{\chi}_{1}^{0}$ 
and $\widetilde{\ell}_{1}$ is smaller than tau mass
\begin{equation}
	\delta m \equiv m_{\widetilde{\ell}_{1}} - 
	m_{\widetilde{\chi}_{1}^{0}} < m_{\tau} \hspace{2mm}.
\end{equation}
Assuming flavor conservation {\it i.e.} $\widetilde{\ell}_1$ is purely RH stau, open decay channels of 
$\widetilde{\ell}_{1}$ are 
\begin{equation}
\begin{split}
	&\widetilde{\ell}_{1} \to \widetilde{\chi}_{1}^{0}  \nu_{\tau}  \pi, ~ 
	\widetilde{\ell}_{1} \to \widetilde{\chi}_{1}^{0}  \nu_{\tau}  a_{1}, ~ 
	\widetilde{\ell}_{1} \to \widetilde{\chi}_{1}^{0}  \nu_{\tau}  \rho, 
	\\
	&\widetilde{\ell}_{1} \to \widetilde{\chi}_{1}^{0}  \nu_{\tau} 
	\ell \bar{\nu}_{\ell} \ \ (\ell \ni e, \mu)\,,  
\label{Eq:many-body}
\end{split}     
\end{equation}
where $\pi$, $a_1$ and $\rho$ are light mesons.
Due to the phase space suppression and higher order coupling 
the $\widetilde{\ell}_{1}$ becomes a long-lived 
particle~\cite{Profumo:2004qt, Jittoh:2005pq}. 
If the lepton flavor is violated, the following 2-body decays are allowed, 
\begin{equation}
\begin{split}
	\widetilde{\ell}_{1} \to \widetilde{\chi}_{1}^{0} 
	\ell \ \ (\ell \ni e, \mu) \hspace{2mm}.  
\label{Eq:LFV2body}	
\end{split}     
\end{equation}
In fact the longevity depends on the degeneracy in mass and also 
on the magnitude of lepton flavor violation~\cite{Kaneko:2008re, Kaneko:2011qi}. 
As we will see in section \ref{sec:bbn}, we have to assume 
$\delta m<m_\mu$, so the main decay mode is the 2-body 
decay $\widetilde{\ell}_1\to\widetilde{\chi}_1^0 e$ and 
therefore the lifetime of the slepton $\tau_{\tilde{l}}$ is given by 
\begin{equation}
\label{eq:lifetime}
\tau_{\tilde{l}}(\tilde{l}_1\to \tilde{\chi}_{1}^{0}+e) \simeq
\frac{8\pi}{g^2 \tan^2{\theta_{W}}}
\frac{m_{\tilde{l}}}{(\delta m)^2} 
\frac{1}{\cos^2{\theta_{e}}+4\sin^2{\theta_{e}}}\frac{1}{C_{e}^2}
\end{equation}
up to leading order of $(\delta m)^2$, where $g$ is the gauge coupling
of $SU(2)$ and $\theta_W$ is the Weinberg angle, respectively.

The long-lived $\widetilde{\ell}_{1}$ has significant effect 
on light element abundances through exotic nuclear processes 
in the BBN era. 
To quantitatively determine this effect, we evaluate the number 
density of $\widetilde{\ell}_{1}$ on the era. As we will see, 
it is closely related with the relic density of 
$\widetilde{\chi}_{1}^{0}$ and it depends on not only 
$\delta m$ but also on the magnitude of lepton flavor 
violation.
Here we take decoupling limit of SUSY particles except for 
$\widetilde{\chi}_{1}^{0}$ and $\widetilde{\ell}_{1}$.

\subsubsection{Dark matter relic density} 
\label{subsec:dm}

After SUSY particles ($\widetilde{\chi}_{1}^{0}$ and 
$\widetilde{\ell}_{1}$) are chemically decoupled from SM 
sectors, their total density, $n = n_{\widetilde{\chi}_{1}^{0}} 
+ n_{\widetilde{\ell}_{1}^{-}} + n_{\widetilde{\ell}_{1}^{+}}$, 
will be frozen. 
Since all of SUSY particles eventually decays into the LSP
$\widetilde{\chi}_{1}^{0}$, so that the dark matter relic density is
indeed the total density.
We find Boltzmann equation of the total density by adding 
each one of $n_{\widetilde{\chi}_{1}^{0}}$ and 
$n_{\widetilde{\ell}_{1}^{\pm}}$~\cite{Griest:1990kh, 
Edsjo:1997bg}, 
\begin{equation}
\begin{split}
	\frac{dY_{n}}{dz} 
	= 
	\frac{-s}{Hz} 
	\sum_{i,j=\widetilde{\chi}_{1}^{0}, \widetilde{\ell}_{1}^{\pm}} 
	\langle \sigma v \rangle_{ij \to \text{SM}}
	\Bigl[ Y_{i} Y_{j} - Y_{i}^{eq} Y_{j}^{eq}  \Bigr],  
\end{split}     
\end{equation}
where $z=m_{\widetilde{\chi}_{1}^{0}}/T$, $Y_{i} = 
n_{i}/s$ is the number density of a species $i$ 
normalized to the entropy density $s$, and $Y_{n} = n/s$, respectively.
Here $H$ denotes the Hubble expansion rate, $\langle \sigma v \rangle_{ij \to \text{SM}}$ represents 
thermally averaged cross-section for an annihilation 
channel $i j \to \text{SM particles}$. 
Relevant processes and the cross-sections are given in 
Ref.~\cite{Nihei:2002sc}. 
We search for favored parameters by numerically solving 
the equation to fit $n$ to the observed dark matter 
density~\cite{Patrignani:2016xqp} 
\begin{equation}
	0.1126 \leq \frac{m_{\widetilde{\chi}_{1}^{0}} \, n h^{2}}
	{\rho_{c}} \leq 0.1246 ~~ (3\sigma \textrm{ C.L.}) \hspace{2mm}, 
\end{equation}
where $h=0.678$ is the Hubble constant normalized to $H_{0} = 
100\,\text{km\,s}^{-1}\,\text{Mpc}^{-1}$, and $\rho_{c} 
= 1.054 \times 10^{-5}\,\text{GeV\,cm}^{-3}$ is the critical 
density of the universe. 



\subsubsection{Number density of long-lived slepton} 
\label{subsec:nslepton}

Even after the chemical decoupling, although the total density 
remains the current dark matter density, the ratio 
of each number density of $\widetilde{\chi}_{1}^{0}$, 
$\widetilde{\ell}_{1}^{-}$, and $\widetilde{\ell}_{1}^{+}$ 
continues to evolve. 
As long as the kinetic equilibrium with the SM sector is 
maintained, $\widetilde{\ell}_{1}$ and $\widetilde{\chi}_{1}^{0}$
follow the Boltzmann distribution, and hence 
$\widetilde{\ell}_{1}^{-}$ number density until the kinetic 
decoupling is 
\begin{equation}
\begin{split}
	n_{\widetilde{\ell}_{1}^{-}} 
	&= 
	\frac{n_{\widetilde{\ell}_{1}^{-}}}{n_{\widetilde{\chi}_{1}^{0}}} 
	\frac{n_{\widetilde{\chi}_{1}^{0}}}{n} n 
	=e^{-\delta m/T} 
	\frac{n}{2 \left( 1+e^{-\delta m/T} \right)}\,. 
\label{Eq:nslepton}	
\end{split}     
\end{equation}
We focus on the parameter space where $\delta m < m_{\mu}$, $m_\mu$ being the muon mass.
Then the lifetime of $\widetilde{\ell}_{1}$ is long enough, and 
we are able to solve the $^7$Li and $^6$Li 
problems~\cite{Jittoh:2008eq, Kohri:2012gc}.
Processes maintaining the kinetic equilibrium in the space 
are\footnote{Note that the process $\widetilde{\ell}_{1}^{\pm} 
\gamma \leftrightarrow \widetilde{\chi}_{1}^{0} e^{\pm}$ 
must not be included. The process should be incorporated 
into a corrective part of the decay (inverse decay) 
$\widetilde{\ell}_{1}^{\pm} \leftrightarrow 
\widetilde{\chi}_{1}^{0} e^{\pm}$. 
Similarly, if the decay $\widetilde{\ell}_{1}^{\pm} 
\leftrightarrow \widetilde{\chi}_{1}^{0} \mu^{\pm}$ is open, 
the process $\widetilde{\ell}_{1}^{\pm} \gamma \leftrightarrow 
\widetilde{\chi}_{1}^{0} \mu^{\pm}$ also must not be taken into account.} 
\begin{equation}
\begin{split}
	&
	\widetilde{\ell}_{1}^{\pm} \gamma \leftrightarrow 
	\widetilde{\chi}_{1}^{0} \tau^{\pm}, ~~ 
	\widetilde{\ell}_{1}^{\pm} \gamma \leftrightarrow 
	\widetilde{\chi}_{1}^{0} \mu^{\pm}, 
	\\&
	\widetilde{\ell}_{1}^{\pm} \tau^{\mp} \leftrightarrow 
	\widetilde{\chi}_{1}^{0} \gamma, ~~ 
	\widetilde{\ell}_{1}^{\pm} \mu^{\mp} \leftrightarrow 
	\widetilde{\chi}_{1}^{0} \gamma, ~~ 
	\widetilde{\ell}_{1}^{\pm} e^{\mp} \leftrightarrow 
	\widetilde{\chi}_{1}^{0} \gamma. 
\label{Eq:RatioChange}	
\end{split}     
\end{equation}
Even for a tiny lepton flavor violation (LFV), flavor changing processes are relevant 
due to much larger densities of $e$ and $\mu$ compared 
with that of $\tau$ for the universe temperature smaller than 
$m_{\tau}$. 
For example, for a reference universe temperature 
$T=70$\,MeV, reaction rates of these processes are 
\begin{equation}
\begin{split}
	\frac{\langle \sigma'v \rangle_{\widetilde{\ell}_{1} e 
	\leftrightarrow \widetilde{\chi}_{1}^{0} \gamma} n_{e}}
	{\langle \sigma'v \rangle_{\widetilde{\ell}_{1} \tau 
	\leftrightarrow \widetilde{\chi}_{1}^{0} \gamma} n_{\tau}} 
	\simeq  
	\left( 1.08 \times 10^{9} \right) C_{e}^{2}\hspace{2mm},
\label{Eq:RatioRate1}	
\end{split}     
\end{equation}
\begin{equation}
\begin{split}
	\frac{\langle \sigma'v \rangle_{\widetilde{\ell}_{1} \mu 
	\leftrightarrow \widetilde{\chi}_{1}^{0} \gamma} n_{\mu}}
	{\langle \sigma'v \rangle_{\widetilde{\ell}_{1} \tau 
	\leftrightarrow \widetilde{\chi}_{1}^{0} \gamma} n_{\tau}} 
	\simeq  
	\left( 9.93 \times 10^{7} \right) C_{\mu}^{2} \hspace{2mm}.
\label{Eq:RatioRate2}	
\end{split}     
\end{equation}
Here $\sigma'$ represents the cross-section of relevant processes 
for kinetic equilibrium. 
As long as $C_{e} \gtrsim 3.2 \times 10^{-5}$ and
$C_{\mu} \gtrsim 1.0 \times 10^{-4}$, flavor changing processes
maintain the kinetic equilibrium, and hence reduce
$n_{\widetilde{\ell}_{1}^{-}}$. This means that such a small flavor mixing can 
decrease $n_{\widetilde{\ell}_{1}^{-}}$ significantly.

The kinetic decoupling is determined by solving coupled 
Boltzmann equations for $\widetilde{\chi}_{1}^{0}$, 
$\widetilde{\ell}_{1}^{-}$, and $\widetilde{\ell}_{1}^{+}$ with 
the initial condition Eq.~\eqref{Eq:nslepton}~\cite{Jittoh:2010wh}, 
\begin{equation}
\begin{split}
	\frac{dY_{\widetilde{\chi}}}{dz} 
	= 
	&- \frac{1}{Hz} 
	\sum_{i \neq \widetilde{\chi}_{1}^{0}}
	\biggl\{ 
	s 
	\langle \sigma'v \rangle_{\widetilde{\chi}_{1}^{0} X \leftrightarrow iY}
	\bigl[ Y_{\widetilde{\chi}} Y_{X}^{eq} - Y_{i} Y_{Y}^{eq} \bigr] 
	\\&  + 
	\langle \Gamma \rangle \, 
	\bigl[ Y_{\widetilde{\chi}_{1}^{0}} \bigl( sY_{X}^{eq} \bigr) 
	\bigl( sY_{X}^{eq} \bigr) ... - Y_{i}\bigr]
	\biggr\}, 
\end{split} 
\label{eq:abandanceChi}    
\end{equation}
\begin{equation}
\begin{split}
	\frac{dY_{\widetilde{\ell}_{1}^{\pm}}}{dz} 
	= 
	&- \frac{1}{Hz} 
	\sum_{i \neq \widetilde{\ell}_{1}^{\pm}}
	\biggl\{ 
	s 
	\langle \sigma'v \rangle_{\widetilde{\ell}_{1}^{\pm} X \leftrightarrow iY}
	\bigl[ Y_{\widetilde{\ell}_{1}^{\pm}} Y_{X}^{eq} - Y_{i} Y_{Y}^{eq} \bigr] 
	\\& + 
	\langle \Gamma \rangle \, 
	\bigl[ Y_{\widetilde{\ell}_{1}^{\pm}} - Y_{\widetilde{\chi}_{1}^{0}} 
	\bigl( sY_{X}^{eq} \bigr) \bigl( sY_{X}^{eq} \bigr) ... \bigr]
	\biggr\}. 
\end{split}     
\label{eq:abandanceSlepton}    
\end{equation}
Here $\Gamma$ represents $\widetilde{\ell}_{1}$ decay rate of channels 
in Eqs.~\eqref{Eq:many-body} and \eqref{Eq:LFV2body}.

\subsection{Big-Bang Nucleosynthesis}\label{sec:bbn}

To solve the Lithium problem(s), we need a long-lived particle 
so that it survives until BBN starts,
more precisely synthesis of $^7$Be begins. Fortunately, our model
does have such a long-lived particle, \mbox{\it i.e.}, $\widetilde{\ell}_1$. 
This slepton can effectively destruct $^7$Be which would be 
$^7$Li just after the BBN era. Since at the BBN
era would-be $^7$Li exists as $^7$Be, destructing $^7$Be effectively
means reducing $^7$Li primordial abundance. This long-lived slepton
with degenerate mass can offer the solution to the $^7$Li
problem~\cite{Jittoh:2007fr, Jittoh:2008eq, Jittoh:2010wh,
Jittoh:2011ni, Kohri:2012gc, Konishi:2013gda, Kohri:2014jfa,
Kusakabe:2010cb,Cyburt:2012kp,Kusakabe:2013tra, Kusakabe:2014moa, Yamazaki:2014fja}.
In addition, several articles~\cite{Asplund:2005yt, Coc:2014oia, Mukhamedzhanov:2016ecq} report that 
there are significant amount of $^6$Li though the standard BBN cannnot predict $^6$Li abundance.

Since we add the RH Majorana neutrinos, these Yukawa couplings are the seed of LFV, 
we have another constraint to impose the longevity of the
lifetime.  To ensure the longevity of the lifetime, only a very tiny
electron and muon flavor can mix in the
NLSP~\cite{Jittoh:2005pq,Kohri:2012gc}.  With keeping these facts in
our mind, here we briefly recapitulate how to solve the Lithium problem(s).

\subsubsection{Non-standard nuclear reactions in stau-nucleus bound state}  
\label{Sec:non-standard} 

We have constraints for the parameters at low energy so that BBN with 
the long-lived slepton works well. To see it we have to take into account the followings:
\begin{enumerate}
 \item[(1)]
Number density of the slepton at the BBN era
\item[(2)]
Non-standard BBN process
\begin{enumerate}
 \item[(a)]
Internal Conversion~\cite{Jittoh:2007fr, Bird:2007ge}
\item[(b)]
Spallation~\cite{Jittoh:2011ni}
\item[(c)]
Slepton catalyzed fusion~\cite{Pospelov:2006sc}
\end{enumerate}
\end{enumerate}

Number density is calculated by numerically solving
Eqs.~(\ref{eq:abandanceChi}) and (\ref{eq:abandanceSlepton}) if the
lifetime is long enough. From this requirement we obtain a constraint
$C_\mu <\mathcal{O}(10^{-5})$ and $C_e<\mathcal{O}(10^{-7})$ with the assumption
$\delta m<m_\mu$ \cite{Kohri:2012gc}.

In addition, since its lifetime must be long enough ($\ge 1700$ s) there is more
stringent constraint on $C_e$ with $\delta m$ as has pointed out in Ref.~\cite{Kohri:2012gc}.
\begin{equation}
 C_e \delta m < 3.5 \times 10^{-9} ~{\rm MeV ~~ for } \ ~ \sin\theta_e=0.6 \hspace{2mm}.
\label{eq:upperboundOnCe}
\end{equation}


\subsubsection{Non-standard nuclear interactions}

\paragraph{Internal Conversion:}  \label{Sec:Internal}  

In a relatively early stage of the BBN, the long-lived slepton forms a
bound state with $^{7}$Be and $^{7}$Li nucleus respectively. These bound states
give rise to internal conversion processes~\cite{Jittoh:2007fr},
\begin{subequations}
\begin{align}
    (^{7}\text{Be}\,\widetilde{\ell}_1^{-}) 
    &\to 
	\widetilde{\chi}_{1}^{0} + \nu_{\tau} + {}^{7}\text{Li} \hspace{2mm},
    \label{Eq:internal_Be}
    \\
    (^{7}\text{Li}\,\widetilde{\ell}_1^{-}) 
    &\to 
	\widetilde{\chi}_{1}^{0} + \nu_{\tau} + {}^{7}\text{He} \hspace{2mm}.
    \label{Eq:internal_Li}
\end{align}
\label{Eq:internal_BeLi}
\end{subequations}
\\[-4mm]
The daughter $^{7}$Li nucleus in the process
Eq.~\eqref{Eq:internal_Be} is destructed either by an energetic proton
or the process \eqref{Eq:internal_Li} while the daughter $^{7}$He
nucleus in the process Eq.~\eqref{Eq:internal_Li} immediately decays
into $^{6}$He nucleus and neutron, then rapid spallation processes by
the background particles convert the produced $^{6}$He into harmless
nuclei, \mbox{\it e.g.} $^{3}$He, $^{4}$He \mbox{\it etc}.  Hence the
non-standard chain reactions by the long-lived slepton could yield
smaller $^{7}$Be and $^{7}$Li abundances than those in the standard
BBN scenario, that is precisely the requirement for solving the
$^{7}$Li problem. This is the scenario we proposed.

We find that the time scale of the reaction is much shorter than the
BBN time scale as long as $\delta m$ is larger than several MeV. A
parent nucleus is converted into another nucleus immediately once
the bound state is formed.  The bound state formation makes the
interaction between the slepton and a nucleus more efficient by two
reasons: First, the overlap of wave functions of the slepton and a
nucleus becomes large since these are confined in the small
space. Second, the short distance between the slepton and a nucleus
allows virtual exchange of the hadronic current even if
$\delta m < m_{\pi}$.

\paragraph{Non-standard process with bound Helium:}
The slepton forms a bound state with $^4$He as well. This fact causes
two non-standard processes. One of these processes is the spallation
process of the $\mathrm{^{4}He}$ nucleus \cite{Jittoh:2011ni},
\begin{subequations}
\begin{align}
    ({}^{4}\text{He} \,\widetilde{\ell}_1^{-}) 
    &\to 
	\widetilde{\chi}_{1}^{0} + \nu_{\tau} + \text{t} + \text{n} \hspace{2mm},
    \label{eq:spal-tn}
    \\
    ({}^{4}\text{He} \,\widetilde{\ell}_1^{-}) 
    &\to 
	\widetilde{\chi}_{1}^{0} + \nu_{\tau} + \text{d} + \text{n} + \text{n} \hspace{2mm},
    \label{eq:spal-dnn}
	\\ 
	({}^{4}\text{He} \,\widetilde{\ell}_1^{-}) 
	&\to 
	\widetilde{\chi}_{1}^{0} + \nu_{\tau} + \text{p} + \text{n} + \text{n} + \text{n} \hspace{2mm}.
	\label{eq:spal-pnnn}
\end{align}
	\label{Eq:spa}
\end{subequations}
\\[-4mm]
and the other channel is called slepton-catalyzed fusion \cite{Pospelov:2006sc}; 
\begin{equation}
	(\mathrm{^{4}He} \,\widetilde{\ell}_1^{-}) + \mathrm{d}
	\to
	\widetilde{\ell}_1^- + \mathrm{^{6}Li}  \hspace{2mm}.
	\label{Eq:catalyzed}
\end{equation}
Since the LFV coupling and $\delta m$ determines which light elements
are over-produced by these non-standard reactions, 
we need careful study of the evolution of the slepton-${}^{4}\text{He}$ 
bound state for the parameter space of $C_\alpha$'s and $\delta m$.
In general the spallation
process is disastrous. In order to suppress it $\delta m <30$~MeV must
be fulfilled.

The catalyzed fusion process enhances the $^6$Li
production~\cite{Pospelov:2006sc}.  Thermal averaged cross-section of
the catalyzed fusion is precisely calculated in
Refs.~\cite{Hamaguchi:2007mp, Kamimura:2008fx}, which is much larger
than that of the $^6$Li production in the Standard BBN,
$^{4}\text{He} + d \to {}^{6}\text{Li} + \gamma$, by 6-7 orders of
magnitude. The over-production of $^{6}$Li nucleus by the catalyzed
fusion process leads stringent constraints on $(\delta m)^2 C_e^2$
from below to make the slepton lifetime shorter than 5000
s~\cite{Kohri:2012gc}.  With the lower bound on the lifetime 1700 s,
in addition to the upper bound on $C_e$, Eq.~(\ref{eq:upperboundOnCe})
we have lower bound on it. For $\delta m =10$~MeV and
$\sin\theta_e=0.6$,
\begin{equation}
 1700~{\rm s} \le \tau_{\widetilde{\ell}} \le 5000~{\rm s} \Leftrightarrow
  2.0\times10^{-10}\le C_e\le 3.5\times 10^{-10}
\label{eq:reqForCe7Li}
\end{equation}
is required.

Furthermore there are several reports~\cite{Asplund:2005yt} that insists
there are significant amount of $^6$Li. If we take it seriously, we can
make use of the catalyzed fusion here and in this case the slepton
lifetime must be between \mbox{3500~s} and \mbox{5000~s} and it corresponds to the requirement
\begin{equation}
 3500~{\rm s} \le \tau_{\widetilde{\ell}} \le 5000~{\rm s} \Leftrightarrow
  2.0\times10^{-10}\le C_e\le 2.5\times 10^{-10}.
\label{eq:reqForCe67Li}
\end{equation}

\subsection{Leptogenesis} 
\label{Sec:lepto} 

We calculate the lepton asymmetry assuming the RH neutrinos being
hierarchical in mass that is generated by the CP asymmetric reactions
of the lightest RH neutrino $N_{1}$ and its superpartner $\tilde{N}_1$.  Typical parameters for solving
the $^7\text{Li}$ and $^6\text{Li}$ problems are
$M_{1} \sim 10^{10}\,\text{GeV}$ and
$|\lambda_{\alpha 1}| \sim 10^{-3}$. Further, the decay parameter should be 
$K \equiv \Gamma_{N_{1}}/H(M_{1}) \sim \mathcal{O}(1)$ and
$K_{\alpha} \equiv K \cdot \text{BR}(N_{1} \to \ell_{\alpha} \phi)
\sim \mathcal{O}(0.1) \ (\alpha \ni {e, \mu, \tau})$.
Here $H(M_{1})$ is the Hubble parameter at the temperature $T= M_{1}$.
In cases where the Leptogenesis in the strong washout regime takes
place $T \lesssim 10^{12} \,\text{GeV}$ and $K_{\alpha}$ are
comparable with each other, the lepton number of each flavor
separately evolves, and it gives rise to $\mathcal{O}(1)$ corrections
to the final lepton asymmetry with respect to where the flavor effects
are ignored~\cite{Nardi:2005hs, Nardi:2006fx}. As studied in
Refs.~\cite{Fong:2011yx, Ishihara:2015uua} the correction could be
significant in SUSY flavored case.

The lepton asymmetry is calculated by a set of the coupled evolution
equations of the number densities of $N_{1}$, $\widetilde{N}_{1}$, and
lepton numbers of each flavor.
Since the super-equilibration is maintained throughout the 
temperature range we consider~\cite{Chung:2009qs}, the 
equality of asymmetries of each lepton and its scalar partner is also 
maintained, and $Y_{B-L} = 2 \times \bigl( Y_{\Delta_{e}} + 
Y_{\Delta_{\mu}} + Y_{\Delta_{\tau}} \bigr)$ with 
$Y_{\Delta_{\alpha}} = B/3 - L_{\alpha}$.  
In the super-equilibration regime, the primary piece 
of the coupled equations are given as 
follows~\cite{Fong:2010qh}
\begin{equation}
\begin{split}
	\frac{dY_{N_1}}{dz} 
	&= 
	\frac{-z}{sH(M_{1})}
	\left( \frac{Y_{N_1}}{Y_{N_1}^{eq}} -1 \right) 
	\Bigl[ \gamma_{N_1} + \gamma_{N_1}^{s1} \Bigr] \,,
\end{split}     
\end{equation}
\begin{equation}
\begin{split}
	\frac{dY_{\widetilde{N}_+}}{dz} 
	&= 
	\frac{-z}{sH(M_{1})} 
	\left( \frac{Y_{\widetilde{N}_+}}{Y_{\widetilde{N}_1}^{eq}} -2 \right) 
	\Bigl[ \gamma_{\widetilde{N}_1} + \gamma_{\widetilde{N}_1}^{s1} \Bigr] \,, 
\end{split}     
\end{equation}
\begin{equation}
\begin{split}
	\frac{dY_{\Delta_{\widetilde{N}}}}{dz} 
	&= 
	\frac{-z}{sH(M_{1})} 
	\left\{ 
	\frac{Y_{\Delta_{\widetilde{N}}}}{Y_{\widetilde{N}_1}^{eq}} 
	\Bigl[ \gamma_{\widetilde{N}_1} + \gamma_{\widetilde{N}_1}^{s2} \Bigr] 
	- \frac{Y_{\Delta \ell}}{Y_{\ell}^{eq}} 
	\Bigl[ \gamma_{\widetilde{N}_1}^{s3} \Bigr] 
	- \frac{Y_{\Delta H_u}}{Y_{H_u}^{eq}} 
	\Bigl[ \gamma_{\widetilde{N}_1}^{s4} \Bigr]
	\right\} \,,
\end{split}     
\end{equation}
\begin{equation}
\begin{split}
	\frac{dY_{\Delta_{i}}}{dz} 
	=& 
	\frac{-z}{sH(M_{1})} 
	\biggl\{ 
	\varepsilon_{i} \left( \frac{Y_{N_1}}{Y_{N_1}^{eq}} -1 \right) 
	\Bigl[ \gamma_{N_1} + \gamma_{N_1}^{s1} \Bigr] 
	+ \varepsilon_{i} 
	\left( \frac{Y_{\widetilde{N}_+}}{Y_{\widetilde{N}_1}^{eq}} -2 \right) 
	\Bigl[ \gamma_{\widetilde{N}_1} + \gamma_{\widetilde{N}_1}^{s1} \Bigr] 
	\\& 
	- \frac{Y_{\Delta \ell}}{Y_{\ell}^{eq}} 
	\left[ 
	\left( \frac{1}{2} \gamma_{N_1}^{i} + \gamma_{N_1}^{s2} \right) 
	+ \left( \gamma_{\widetilde{N}_1}^{i} + \gamma_{\widetilde{N}_1}^{s5} \right)
	\right] 
	\\& 
	- \frac{Y_{\Delta H_u}}{Y_{H_u}^{eq}} 
	\left[ 
	\left( \frac{1}{2} \gamma_{N_1}^{i} + \gamma_{N_1}^{s3} \right) 
	+ \left( \gamma_{\widetilde{N}_1}^{i} + \gamma_{\widetilde{N}_1}^{s6} \right)
	\right] 
	\biggr\} \,.
\end{split}     
\end{equation}
Here $z = M_{1}/T$. We introduced transformed yield values for 
$\widetilde{N}_1$, $Y_{\widetilde{N}_+} \equiv Y_{\widetilde{N}_1} 
+ Y_{\widetilde{N}_1^{*}}$, and $Y_{\Delta_{\widetilde{N}}} \equiv 
Y_{\widetilde{N}_1} - Y_{\widetilde{N}_1^{*}}$. 
$\gamma_{N_1}$ and $\gamma_{\widetilde{N}_1}$ are thermally 
averaged decay rates of $N_1$ and $\widetilde{N}_1$, respectively. 
$\gamma_{X}^{s \, n} \ (n = 1, 2, 3, ...)$ symbolizes a combination 
of thermally averaged cross-sections, and the explicit one is shown 
in Appendix in Ref.~\cite{Fong:2010qh}. Relevant cross-sections 
are given in Ref.~\cite{Plumacher:1997ru}.
Coefficient $C_{\alpha\beta}^{\ell}$ ($C_{\beta}^{H}$) is a conversion factor from 
the asymmetry of $\ell_{\alpha}$ ($H$) to that of $\ell_{\beta}$, 
$\left( n_{\ell_\alpha} - n_{\bar{\ell}_\alpha} \right)/n_{\ell_\alpha}^{eq} = 
- \sum_{\beta} C_{\alpha\beta}^{\ell} \left( Y_{\Delta_\beta}/Y_{\ell}^{eq} \right)$ 
and $\left( n_{H} - n_{\bar{H}} \right)/n_{H}^{eq} = - \sum_{\beta} 
C_{\beta}^{H} \left( Y_{\Delta_\beta}/Y_{\ell}^{eq} \right)$. The entries 
are determined by constraints among the chemical potentials 
enforced by the equilibrium reactions at the stage where the 
asymmetries are generated, $T \sim M_{1}$. 
In our scenario, $M_{1} \sim 10^{10}\,\text{GeV}$, and 
$C_{\alpha\beta}^{l}$ and $C_{\beta}^{H}$ are~\cite{Fong:2010qh} 
\begin{equation}
\begin{split}
   C_{\alpha\beta}^{l} = \frac{1}{3 \times 2148}
\begin{pmatrix} 
   906 & -120 & -120
   \\
   -75 & 688 & -28 
   \\
   -75 & -28 & 688 
\end{pmatrix}, \ ~~~ 
    C^{H} = \frac{1}{2148} 
\begin{pmatrix} 
   37 & 52 & 52
\end{pmatrix}. 
\end{split}      
\end{equation}

\begin{figure}[t!]
\begin{center}
\includegraphics[width=110mm]{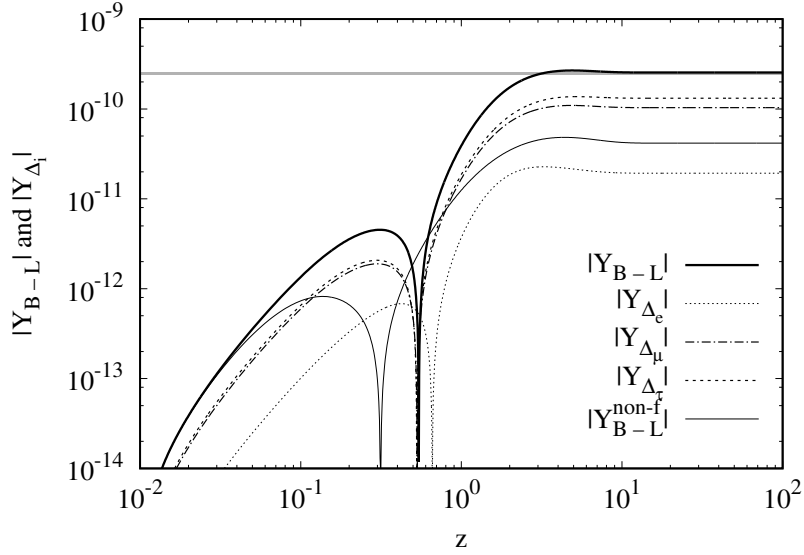}
\caption{Evolutions of $|Y_{B-L}|$ and each lepton asymmetry
  $|Y_{\Delta_{i}}|$ for a typical parameter in this paper.
  Horizontal band (gray) corresponds to the observed baryon
  asymmetry. $|Y_{B-L}^\text{non-f}|$ shows the lepton asymmetry in
  the absence of flavor effect.}
\label{Fig:result1207A}
\end{center}
\end{figure}

The CP asymmetry receives contributions from 
not only the RH neutrinos but also its scalar partner. The 
flavor dependent CP asymmetry for the channel $N_{i} \to 
\ell_{\alpha} \phi$ is defined as 
\begin{equation}\label{eq:epsilon}
\varepsilon^{i}_{\alpha} \equiv 
\frac{\Gamma(N_{i} \to \ell_{\alpha} \phi) - 
\Gamma(N_{i} \to \bar{\ell}_{\alpha} \phi^{\dagger}) }{
 \Gamma(N_{i} \to \ell_{\alpha} \phi) + 
\Gamma(N_{i} \to \bar{\ell}_{\alpha} \phi^{\dagger}) }
\end{equation}
and is obtained as~\cite{Covi:1996wh}, 
\begin{equation}
\begin{split}
	\varepsilon^{i}_{\alpha} 
	= 
	\varepsilon^{i}_{\alpha} (\text{vertex}) + 
	\varepsilon^{i}_{\alpha} (\text{wave}) \hspace{2mm}, 
\label{Eq:CP}	
\end{split}     
\end{equation}
\begin{equation}
\begin{split}
	\varepsilon^{i}_{\alpha} (\text{vertex}) 
	=
	- \frac{1}{8\pi} \sum_{j}
	\frac{M_{j}}{M_{i}}
	\log\left[ 1+ \frac{M_{i}^{2}}{M_{j}^{2}} \right]
	\frac{\Im \left[  
	\left( \lambda^{\dagger} \lambda \right)_{ji} 
	\lambda_{\beta i}^{*} \lambda_{\alpha i} \right]}
	{\left( \lambda^{\dagger} \lambda \right)_{ii}} \hspace{2mm}, 
\label{Eq:vertex}	
\end{split}     
\end{equation}
\begin{equation}
\begin{split}
	\varepsilon^{i}_{\alpha} (\text{wave})
	=
	- \frac{2}{8\pi} \sum_{j} 
	\frac{M_{i}}{M_{j}^{2} - M_{i}^{2}}
	\frac{\Im \left\{ 
	\left[ 
	M_{j} \left( \lambda^{\dagger} \lambda \right)_{ji} 
	+ M_{i} \left( \lambda^{\dagger} \lambda \right)_{ij} 
	\right]
	\lambda_{\beta i}^{*} \lambda_{\alpha i}
	\right\}}
	{\left( \lambda^{\dagger} \lambda \right)_{ii}} \hspace{2mm}. 
\label{Eq:wave}	
\end{split}     
\end{equation}
The CP asymmetries for other channels, 
$N_{i} \to \widetilde{l}_{\alpha} \widetilde{\chi}$, 
$\widetilde{N}_{i} \to l_{\alpha} \widetilde{\chi}$, and 
$\widetilde{N}_{i} \to \widetilde{l}_{\alpha} \phi$, 
are defined similarly, and given as the same results with 
Eqs.~\eqref{Eq:CP}, \eqref{Eq:vertex} and \eqref{Eq:wave}. 

The lepton and slepton asymmetry converts to the baryon 
asymmetry, and the conversion factor in MSSM scenarios is 
$Y_{B} = (8/23) Y_{B-L}$~\cite{Laine:1999wv}. 
The required lepton asymmetry in 3 sigma range is  
\begin{equation}
2.414 \times 10^{-10} 
\lesssim |Y_{B-L}| \lesssim 2.561 \times 10^{-10} \hspace{2mm}
\label{eq:leptonasym} 
\end{equation}
for the observed baryon number $\Omega_{b}h^{2} = 0.0223 \pm 
0.0002$ ($1\sigma$)~\cite{Patrignani:2016xqp}.

Figure~\ref{Fig:result1207A} shows the evolution of lepton 
number for a typical parameter obtained in this study. 
Numerical computations in this work are performed by using 
the complete set of coupled Boltzmann equations. For illustrating 
the importance of flavor effect, we also plot the non-flavored 
result with thin solid line. 
We find $\mathcal{O}(1)$ correction to the final lepton asymmetry 
depending on the presence of the flavor effect. Since 
this correction is introduced into the expected relation between 
$M_{1}$ and $\lambda_{\alpha i}$, the flavor effects are critical 
ingredients to understand the correlation among the BBN, the BAU, 
and the charged LFV in our scenario.

\section{Analysis\label{sec:analysis}}
\subsection{Parameter Space\label{sec:parameterspace}}
\indent  Soft SUSY breaking term in the Lagrangian $\mathscr{L}_{\rm soft}$
contains more than one hundred parameters in general. In order to
perform phenomenological study we make an assumption that three gauge
couplings unified at GUT scale and further for reduction of the number
of parameters. At that scale we presume that there exists a universal
gaugino mass, $m_{1/2}$. Besides, the scalar soft breaking part of the
Lagrangian depends only on a common scalar mass $m_0$ and trilinear
coupling $A_0$, in addition on the ratio of VEVs, $\tan\beta$.  After
fixing a sign-ambiguity in the higgsino mixing parameter $\mu$ we complete five 
SUSY parameter space of the CMSSM:
\begin{equation}
\label{eq:msugraparameterspace}
m_{1/2},\ m_0,\ A_0,\ \tan\beta,\ {\rm sign}(\mu) \hspace{2mm}.
\end{equation}
Note that we have demonstrated our numerical analyses only
in the ${\rm sign}(\mu)>0$ case. 


In the neutrino Yukawa couplings, Eq.\eqref{eq:eq9}, 
there are 18 parameters since the matrix is $3\times 3$ complex matrix.
We use the low-energy observed quantifies \mbox{({\it i})} three LH
neutrino masses $m_{\nu_{1}},m_{\nu_{2}},m_{\nu_{3}}$ \mbox{({\it
    ii})} three mixing angles
$\sin\theta_{23}, \, \sin\theta_{13}, \, \sin\theta_{12}$ in
$U_{\mathrm{MNS}}$ (Eq.\eqref{eq:MNS}) and \mbox{({\it iii})} three
CP-violating phases $\alpha, \beta, \delta$ (Eq.\eqref{eq:expmixings})
as input parameters. They are given in Sec.~\ref{sec:notation}.
There are 9 model parameters, which we express in terms of 3 RH Majorana
neutrino masses $M_1, M_2, M_3$ at GUT scale and remain 3 complex
angles in $R$ matrix. Thus, there are total 9 free parameters and 9
experimentally ``observed'' data in the Dirac Yukawa couplings.
 
The low-energy SUSY spectra and the low-energy flavor observables were
computed by means of the SPheno-3.3.8~\cite{Porod:2003um,Porod:2011nf}
using two-loop beta functions with an option of the precision as
quadrupole because the slepton flavor mixing is required to be
$10^{-12}$ order or even smaller. During these computation we apply
the set of constraints displayed in Table~\ref{tab:flavor}.  We
generate SLHA format files and send them to
micrOMEGAs\_4.3.5~\cite{Belanger:2008sj,Belanger:2010gh,Barducci:2016pcb}
which computes the neutralino relic density $\Omega h^2$ and the
spin-independent scattering cross-section with nucleons, as we will briefly mention below.


\begin{table}[t!]
\centering
\begin{tabular}{|c|c|c|}
\hline
Quantity & & Reference \\\noalign{\hrule height 2pt}
$\Omega h^2$ & $[0.1126, 0.1246]$ & \cite{Patrignani:2016xqp} \\ \hline
$m_h$ & $(124.4,125.8)$ GeV & \cite{Patrignani:2016xqp} \\ \hline
${\rm BR}(B\to s \gamma)$ & $[2.82, 3.29] \times 10^{-4}$ &
\cite{Amhis:2016xyh} \\
${\rm BR}(B_s \rightarrow \mu^{+} \mu^{-}) $ & $2.8^{+2.1}_{-1.8} \times 10^{-9}$ &
\cite{CMS:2014xfa} \\
${\rm BR}(B_u \rightarrow \tau \bar{\nu})$ & $0.52 < R_{B \tau \nu} <
2.61$ &\cite{Asner:2010qj} \\
$a_\mu$ & $[1.97, 50.2]\times10^{-10}$ & \cite{Bennett:2006fi} \\ \hline
\end{tabular}
\caption{\label{tab:flavor} The experimental constraints.}
\end{table}


\subsection{Determining input parameters}
\label{subsec:input-param}
In this subsection we discuss in detail how we have investigated very wide range of 
parameter space. 
In principle we must set all the parameters simultaneously so that all
the requirement are fulfilled. However conceptually we can set the parameters
step by step with the small correction from the following steps.
\subsubsection{The CMSSM  parameters}
\label{sec:cmssm}
\begin{table}[t!]
    \begin{tabular}{c}
      \begin{minipage}[t]{.45\textwidth}
\begin{center}
  \begin{tabular}{|c||r|} \hline
    ~Input Parameters~ & ~~value~~~~~~~~  \\ \hline \hline
    $m_{0}$ & 707\,$(\mathrm{GeV})$  \\
    $m_{1/2}$ & 887\,$(\mathrm{GeV})$   \\
    $A_{0}$ & -3089\,$(\mathrm{GeV})$   \\ 
    $\tan{\beta}$ & 25  \\
    $\mu / |\mu|$ & $+1$  \\ \hline
    $m_{\nu_{1L}}$ & $10^{-3} \, (\mathrm{eV})$   \\
    $m_{\nu_{2L}}$ & ~~$4.04 \times 10^{-3} \, (\mathrm{eV})$   \\ 
    $m_{\nu_{3L}}$ & $1.18 \times 10^{-2} \, (\mathrm{eV})$   \\  
    $M_{1}$ & $2.0\times 10^{10}(\mathrm{GeV})$   \\ 
    $M_{2}$ & $8.0\times 10^{10}(\mathrm{GeV})$   \\ 
    $M_{3}$ & $8.0\times 10^{11}(\mathrm{GeV})$   \\ \hline
    $\alpha$ & 0  \\ 
    $\beta$ & 0  \\ 
    $\delta$ & 261  \\ \hline
    $x_{12}$ & 2.28948  \\ 
    $x_{13}$ & 3.56000  \\ 
    $x_{23}$ & 4.80532  \\ 
    $y_{12}$ & 1.02  \\ 
    $y_{13}$ & 0.1  \\ 
    $y_{23}$ & 0.1  \\ \hline\hline 
  \end{tabular}
\end{center}
      \end{minipage}
      \begin{minipage}[t]{.45\textwidth}
\begin{center}
\vspace{-4cm}
  \begin{tabular}{|c||r|} \hline
    ~output parameters~ & value~~~~~~  \\ \hline \hline
    $C_{e}$ & ~~$3.28 \times 10^{-10}$  \\
    $C_{\mu}$ & $2.94 \times10^{-6}$  \\
    $\sin{\theta_{e}}$ & 0.188  \\
    $\tau_{\widetilde{l}}$ & 4217\,(s)  \\[2mm] 
\hline\hline
  \end{tabular} \\[2cm]

  \begin{tabular}{|c||r|} \hline
    ~output parameters~ & value~~~~~~~~~  \\ \hline \hline
    $\Omega h^{2}$ & 0.115  \\
    $m_{\widetilde{\chi}_1^0}$ & 379.6\,(GeV)  \\
    $\delta m$ & ~~$1.01 \times 10^{-2}$\,(GeV)  \\ \hline\hline 
  \end{tabular}
  \end{center}
     \end{minipage}
    \end{tabular}
 \caption{The example of input parameters and output parameters.}
\label{tab:anexample} 
\end{table}

Let us start with the constraints on the lightest neutralino mass from
relic abundance.  For our analysis we take into account cosmological
data -- dark matter abundance -- that arises from the Planck satellite
analysis~\cite{Ade:2015xua}. In this article we request the neutralino
relic density, $\Omega h^2$, must satisfy the $3$ sigma range:
$\Omega h^2\in[0.1126, 0.1246]$~\cite{Patrignani:2016xqp}.  In CMSSM
type theory the lightest neutralino mass will be of order of 400
GeV. In the framework of MSSMRN which we consider the lightest
neutralino mass becomes about 380 GeV. What is more we fix the mass
difference $\delta m=0.01$~GeV as already studied~\cite{Kohri:2012gc},
furthermore we decide to use $\tan\beta=25$ because with this value we
can easily obtain a right amount of the relic density which must be
within $3$ sigma rage of cosmological data. Accordingly, three of SUSY
parameters, $m_{1/2}$, $A_0$ and $\tan\beta$, we set in the following values
\begin{equation}
m_{1/2} = 887.0\,(\mathrm{GeV})  \hspace{2mm}, 
\quad  A_{0}=-3090\,(\mathrm{GeV}) \hspace{2mm},
\quad  \tan\beta=25 \hspace{2mm}.
\end{equation}

At this moment, four SUSY parameters are fixed including sign of the
mu-term, the remaining parameter, the universal scalar mass, $m_0$,
must lie on
\begin{equation}
m_{0} \approx [707.3, 707.4]\,(\mathrm{GeV}) \hspace{2mm}, 
\end{equation}
depending on the mass hierarchy structure of the RH Majorana neutrino
sector for fixing the value of $\delta m=0.01$~GeV\@, not only due to
the logalismical corrections of the corresponding scales but also the
slepton mass running effect which are caused by the Dirac
Yukawa beta-function. However, the effect of Dirac Yukawa runnings for
calculation of the dark matter relic density is negligible and thus we
can safely ignore this effect.

It is important to mention that with the above given values of SUSY
parameters we obtain the SM-like Higgs
mass about 125 GeV, \mbox{\it i.e.} the ``right'' combination of
the values of $\tan\beta$, $A_{0}$ and stop mass generated by the
universal scalar mass $m_0$ are selected in our calculation processes.

We show here an example parameter in the left-panel of Table~\ref{tab:anexample}.
With this parameter set, the flavor mixing $C_{e}$,
$C_{\mu}$, and the mixing angle $\sin{\theta_{e}}$, and 
the lifetime of slepton are calculated. The results are listed in 
the right-top-panel of Table~\ref{tab:anexample}. 
It is clear that our model with these parameters
solve also both $^6$Li and $^7$Li problems.  Furthermore, we have
calculated the observed quantities, the relic density of dark matter,
mass of dark matter, and the mass difference between the NLSP and LSP,
$\delta m$, with same parameters.  The results are displaced in the right-down-panel of 
Table~\ref{tab:anexample}. As has been noted, our result, $\Omega h^2= 0.1154$, satisfies 
the relic density obtained from the Planck satellite analysis~\cite{Ade:2015xua}.


\subsubsection{The Yukawa coupling}
\label{sec:yukawa}

In order to find a set of parameters with which our model -- MSSMRH with
boundary condition at the GUT scale -- we have performed parameter scan in
the following ``systematic'' way. Essentially we do not scan all mass
range of the RH Majorana neutrinos, but we fix the mass
ratio of these particles. It means that the second heaviest and
the heaviest RH Majorana masses are given by a function of the lightest
RH one, hereby we fix the ratio $M_3/M_1 =
40$, and we investigate the following three scenarios in this
article. Namely,
 \begin{enumerate}
\centering
 \item $M_{2}=2\times M_{1}$, $M_{3}=40\times M_{1}$
 \item $M_{2}=4\times M_{1}$, $M_{3}=40\times M_{1}$
 \item $M_{2}=10\times M_{1}$, $M_{3}=40\times M_{1}$
\end{enumerate}
\mbox{\it i.e.} only the ratios of $M_{2}/M_1$ are different in each setup.

Fixing the mass of the lightest RH Majorana neutrino and
arranging the elements of the complex orthogonal matrix $R$, \mbox{\it
  i.e.} real part of the complex angles ($x_{12}$, $x_{13}$, $x_{23}$) 
and complex part of ones, $y_{12}, ~y_{13},~y_{23}$, 
we are now able to calculate the baryon asymmetry. For simplicity
we fix the values of $y_{23}=y_{13}=0.1$ and vary only $y_{12}$ in
the complex part of the mixing angles. 
The real part of complex angles, $x_{12}$, $x_{13}$,
$x_{23}$, are obtained through the electron mixing in slepton
mass matrix, $C_{e}$. To have a enough lifetime of slepton for
solving the Lithium problem, only extremely narrow ranges of $x_{12}$,
$x_{13}$, $x_{23}$, -- of the order of
$10^{-5}$ -- are allowed because of $C_{e}$ being of the order of
$10^{-10}$. 
\begin{figure}[t]
\begin{center}
\includegraphics[width=10cm]{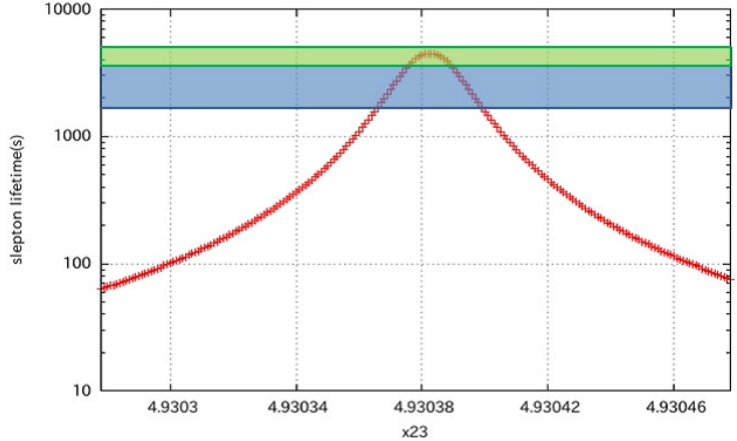}
\caption{The lightest slepton lifetime as a function of $x_{23}$.
The blue and green band corresponds to the lifetime required to solve the $^7$Li problem only 
and both the $^7$Li and $^6$Li problems, respectively.}
\label{fig:lifetime-vs-x23}
\end{center}
\end{figure}
To illustrate how the real parts of the flavor mixing are determined, we show the lifetime of the slepton in 
terms of $x_{23}$ in Figure \ref{fig:lifetime-vs-x23}. The RH Majorana mass is taken to 
$M_1=2.0 \times 10^{10}$ GeV in case 2.  The blue and green bands represent the slepton lifetime 
required to solve only the $^7$Li problem, Eq.\eqref{eq:reqForCe7Li},
and both $^7$Li and $^6$Li problems, Eq.\eqref{eq:reqForCe67Li}, respectively. 
The lifetime changes two orders of magnitude for the narrow range of $x_{23}$ of 
order $10^{-5}$. This is because $C_e \sim 10^{-10}$ is realized due to the fine-tuned cancellation among 
the LFV terms in renormalization group equation running. When $x_{23}$ differs from this range, $C_e$ 
is $\mathcal{O}(10^{-5})$ and hence the slepton lifetime becomes much shorter. 
One can see that the real part $x_{23}$ is determined almost uniquely to solve the Li problems.
No need to say that allowed regions of the real part of
the complex angles are also depend on the mass structure of the RH Majorana
neutrinos thus we have to seek an other tiny parameter space when we
change the value of $M_1$.
Furthermore we check the parameters obtained in this way whether
they reproduce the right amount of baryon asymmetry
as explained in Sec.~\ref{Sec:lepto}.

\subsubsection{The allowed mass region of the lightest right-handed
  Majorana neutrino}
\label{sec:RHmass}

We describe our main results in this subsection. First we discuss the
allowed mass range of the lightest RH Majorana neutrino. We
have found the upper- and lower-limit for mass of the lightest
RH Majorana neutrino corresponding to its hierarchical 
structure. With respect to the Lithium problem, three cases which we have
investigated are listed here:
\begin{enumerate}
 \item case of $M_{2}=2\times M_{1}$, $M_{3}=40\times M_{1}$
\begin{itemize}
\item  Taking into account $^{6}$Li and $^{7}$Li problem 
\begin{equation}
7.8 \times10^{8}  \leq M_{1} \leq 7.0 \times10^{10}~\mathrm{(GeV)} \hspace{2mm}.
\end{equation}

\item Taking into account only $^{7}$Li problem
\begin{equation}
 7.8 \times10^{8} \leq M_{1} \leq 1.0\times10^{11}~\mathrm{(GeV)}\hspace{2mm}.
\end{equation}

\end{itemize}

 \item case of $M_{2}=4\times M_{1}$, $M_{3}=40\times M_{1}$
\begin{itemize}

\item  Taking into account $^{6}$Li and $^{7}$Li problem
\begin{equation}
1.9\times10^{9} \leq M_{1} \leq 7.0\times10^{10}~\mathrm{(GeV)}\hspace{2mm}.
\end{equation}

\item Taking into account only $^{7}$Li problem
\begin{equation}
1.9\times10^{9} \leq M_{1} \leq 1.0 \times10^{11}~\mathrm{(GeV)}\hspace{2mm}.
\end{equation}

\end{itemize}

 \item case of $M_{2}=10\times M_{1}$, $M_{3}=40\times M_{1}$
\begin{itemize}

\item  Taking into account $^{6}$Li and $^{7}$Li problem
\begin{equation}
2.35 \times10^{9}~\mathrm{(GeV)} \leq M_{1}  \hspace{2mm}.
\end{equation}

\end{itemize}
\end{enumerate}
One might be wonder why we do not write the upper limit of the
lightest RH Majorana neutrino mass in the third case. Essentially we do
not need to get the values that definitely exist because the region
where the upper limit would be is already excluded by the current
experiment date of ${\rm BR}(\mu\to e\gamma)$. 

\begin{figure}[t]
\begin{center}
\begin{tabular}{cc}
\includegraphics[width=8.5cm]{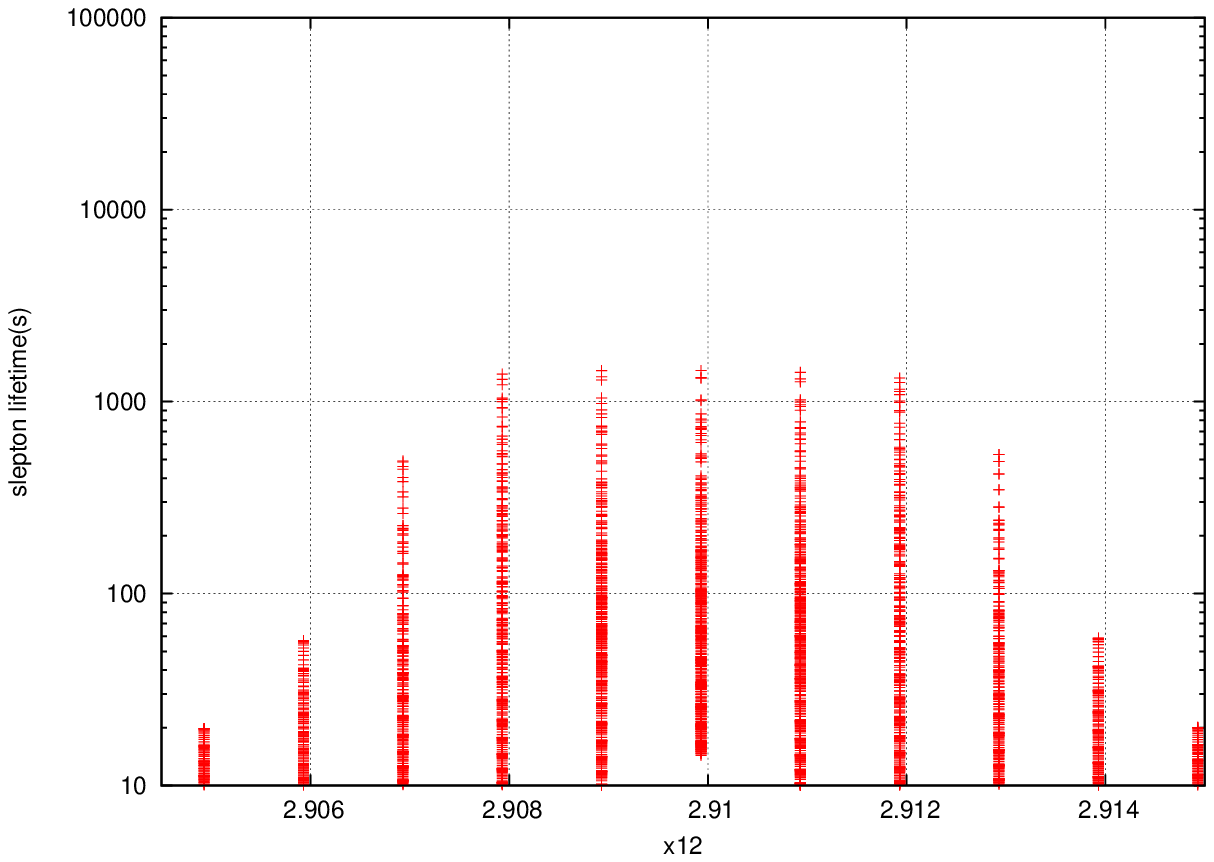}
&
\includegraphics[width=8.5cm]{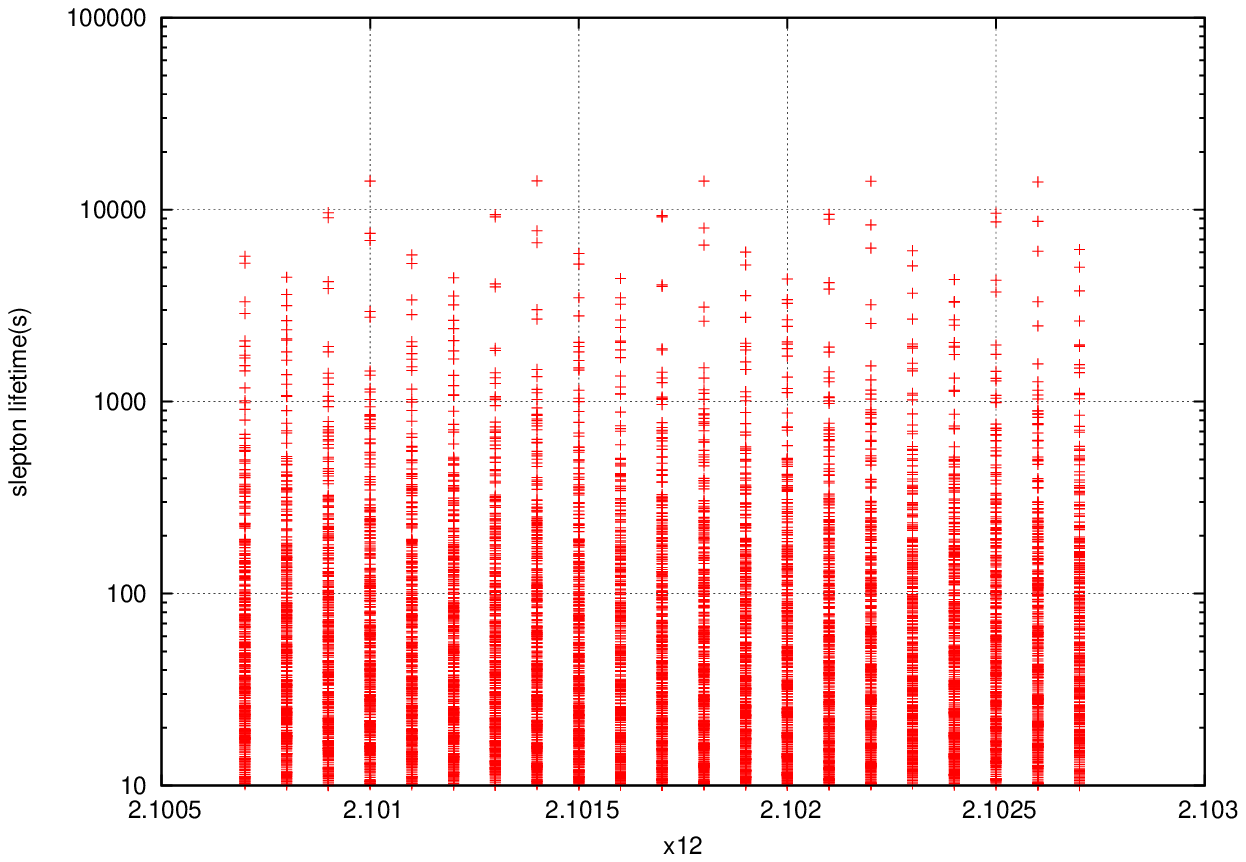} \\
\includegraphics[width=8.5cm]{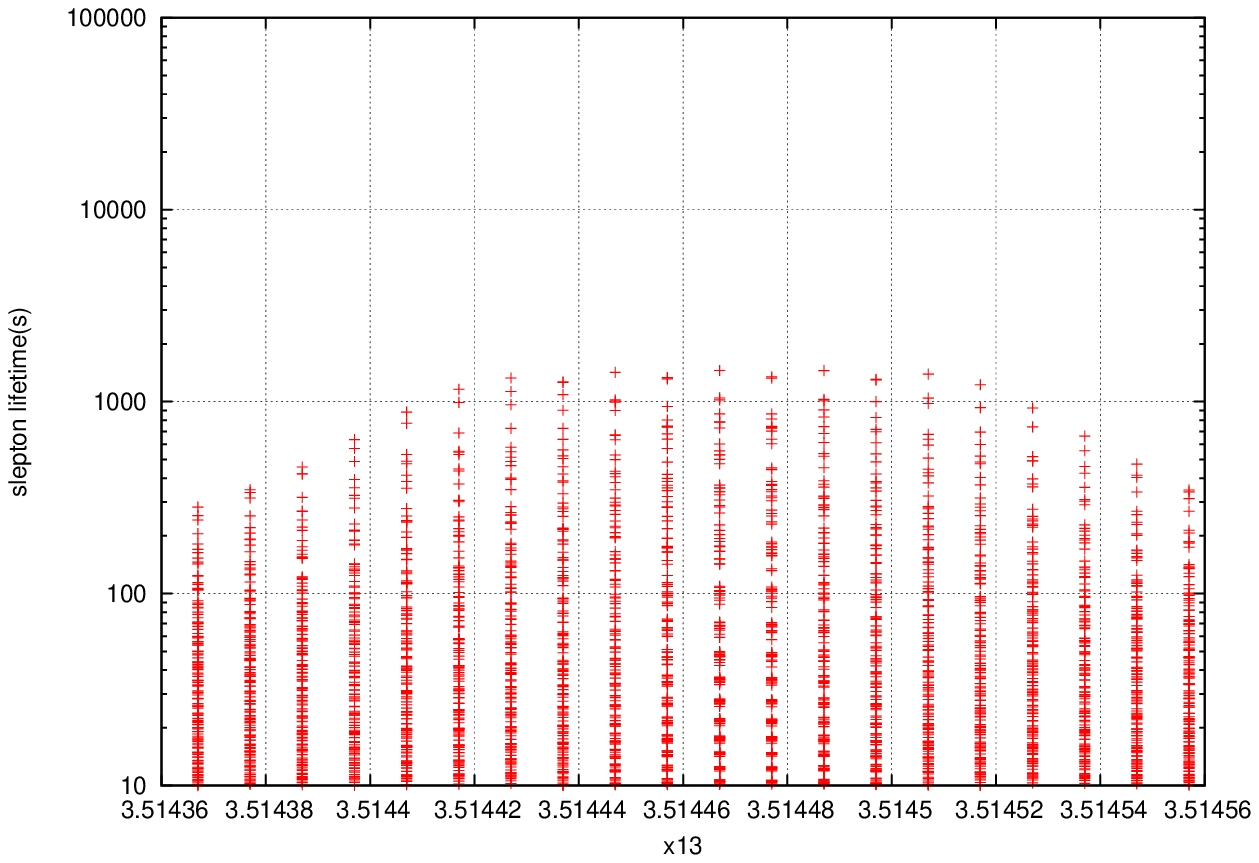}
&
\includegraphics[width=8.5cm]{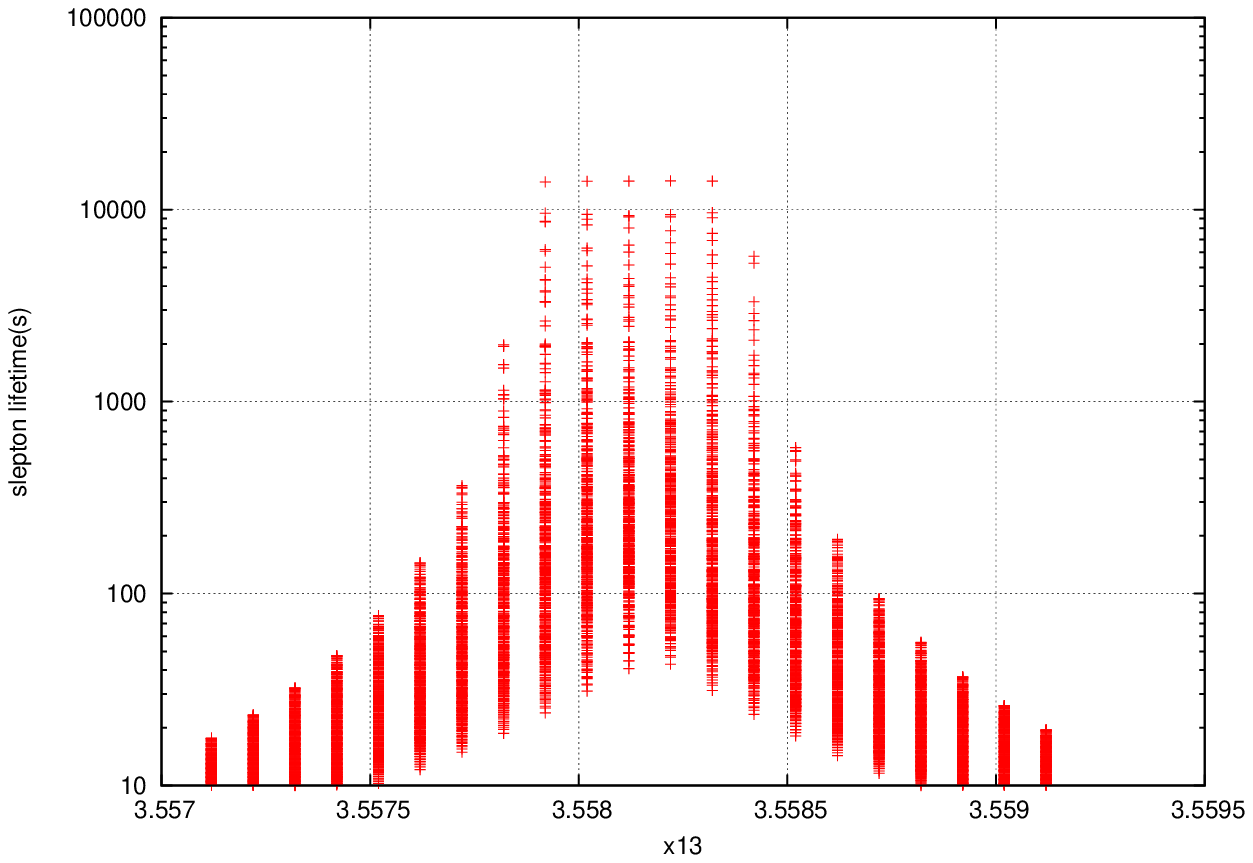} \\
\includegraphics[width=8.5cm]{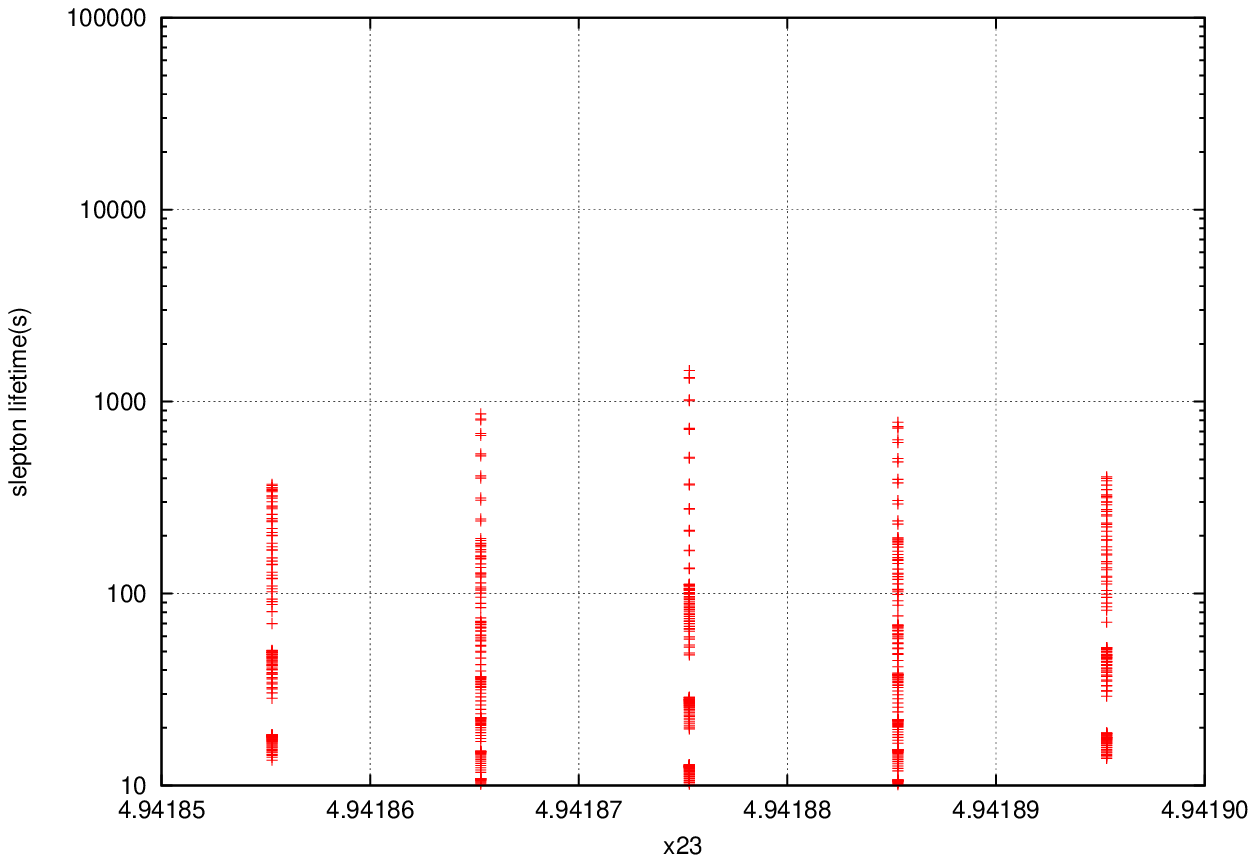}
&
\includegraphics[width=8.5cm]{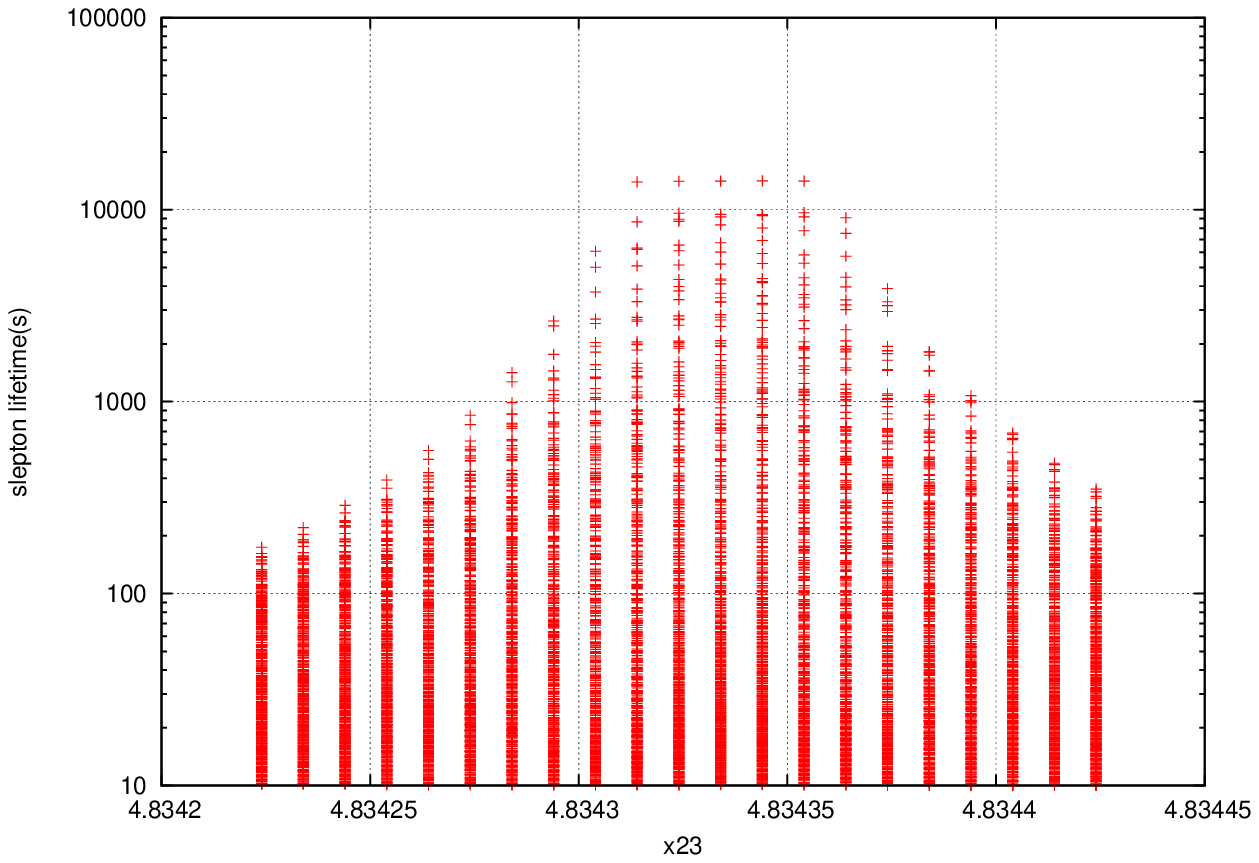} 
\end{tabular}
\end{center}
\caption{The slepton lifetime in terms of $x_{12},~x_{13},~x_{23}$ in case $2$.
In the left and right panel, $M_1$ is taken to $1.2 \times 10^{11}$ GeV and $3.0 \times 10^{10}$ GeV, 
respectively. 
}
\label{fig:lifetime-vs-x2}
\end{figure}
The upper limits of the lightest RH Majoarana neutrino in three
different cases are obtained from the limits of $C_e$ and $C_\mu$,
in fact we need to suppress both the slepton mixing $C_e$ and $C_\mu$.  
Naively, these flavor mixing are scaled to the
Yukawa couplings, so that it is easy to understand why we let the
absolute values of the Dirac Yukawa couplings be
$\left|\lambda _{\alpha i} \right| \ll 1$ to satisfy the experimental
constraints of $C_e$ and $C_\mu$. At the same time, of course we must
satisfy the low-energy neutrino experiment data namely $\Delta m^2$
and three mixing angles according to Eq.~(\ref{mnuKN}) in which we do
not consider an extreme fine-tuning in matrix
multiplications. Therefore the eigenvalues of $M_R$, or the lightest
RH Majorana neutrino mass since we fix the mass ratios $M_2/M_1$ and
$M_3/M_1$, should be lighter than the case of
$\left|\lambda _{\alpha i} \right| \sim 1$.  Further Yukawa coupling
constants square is scaled to the RH Majorana neutrino masses, at a
certain mass it becomes impossible for $C$'s to be small enough. Thus
we have the upper bound for $M_1$. 
In Figures \ref{fig:lifetime-vs-x2}, we show the slepton lifetime as a function of $x_{12},x_{13},x_{23}$ 
to illustrate this explanation. The RH Majorana mass $M_1$ is taken to $1.2 \times 10^{11}$ GeV and 
$3.0 \times 10^{10}$ GeV in the left and right panels, respectively. In the left panels, the lifetime cannot reach 
 $1700$ s even when $x_{12,13,23}$ are fine-tuned.
On the other hand, in the right panels where $M_1$ is taken to be smaller, the lifetime can be  
longer than $1700$ s. Note that all of the real parts are determined in very narrow 
range as we explained in Fig.~\ref{fig:lifetime-vs-x23}.
The flavor mixing $C_e$ is more tightly constrained to
solve the $^6$Li problem (or evade $^{6}$Li over production) and hence
the upper bound is more stringent. 
It is worth noting that with similar reason, we cannot have a
degenerate solution for left-handed neutrino mass since in this case
also rather large Yukawa coupling is necessary.

On the other hand, to reproduce the matter-antimatter asymmetry
generated by Leptogenesis so that the CP violating of Majorana decay
processes $\varepsilon^{i}_{\alpha}$ (Eq.\eqref{eq:epsilon}) should be
large than $10^{-6}$, if one does not take into account the flavor
effects neither does not consider also an accidental fine-tuning
cancellation in $\lambda_\nu^\dagger \lambda_\nu$. Thus we need a sufficiently
large Yukawa couplings and large value of $M_1$ is required (as we
know non-flavor Leptogenesis case, $M_1 \gtrsim 10^9$ GeV
\cite{Davidson:2002qv}.) As it is scaled to the RH Majorana neutrino mass at a
certain point such a sufficiently large coupling can not be realized.

In concluding, we mention that we have found there exists only such
tiny allowed parameter space for the lightest RH Majorana neutrino
where all experimental data and constrains are fulfilled within 3 sigma
range.

\section{Predictions from parameter search} 
\label{subsec:num}
\subsection{Predictions mainly from CMSSM parameters} 
\label{subsec:SUSY}
As explained in Sec.\ref{sec:cmssm}, CMSSM parameter is almost
determined uniquely from $m_{\widetilde{\chi}^0_1}$, $\delta m$, and
SM Higgs mass. Therefore the dark matter relic abundance, SUSY mass
spectrum, and the contribution to the muon magnetic anomalous moment
$g-2$ are more or less predicted uniquely.

In our analysis, the relic abundance of the neutralino density is
\begin{equation}
\Omega h^2=0.115  \hspace{2mm}.
\end{equation}

For calculation of the spin-independent cross-section with nucleon we
use the following values of the quark form-factors in the nucleon
which are the default values in the micrOMEGAs code
\begin{eqnarray}
\label{eq:scalar}
f^p_{d}=0.033\,, \;\; f^p_{u}=0.023\,, \;\; f^p_{s}=0.26\,, \nonumber\\
f^n_{d}=0.042\,, \;\; f^n_{u}=0.018\,, \;\; f^n_{s}=0.26\,. 
\end{eqnarray}
and we get 
\begin{equation}
  \label{eq:SI}
  \sigma^{\rm SI} = 1.05\times 10^{-47} ~ \textrm{cm}^2 \hspace{2mm},
\end{equation}
so that our dark matter candidate satisfies easily the limit of the
spin-independent cross-section with nucleon reported by the LUX
collaboration~\cite{Akerib:2016vxi}, even including the main
uncertainty from the strange quark coefficient. If we use another set
of quark coefficients (the large corrections to $f^{p/n}_{s}$) can
lead to a shift by a factor $2$-$6$ in the spin-independent
cross-section~\cite{Belanger:2008sj}.

Masses of supersymmetric particles are shown in
Table~\ref{SUSYpectra}. Note that these spectrum are predicted just
above the current experimental limits~\cite{Patrignani:2016xqp}.
\begin{table}[h!] 
  \begin{center}
  \begin{tabular}{|c||r||r|} \hline
    ~~particle~~ & ~~mass (GeV)~~ & ~~~~mixing~~~~~~~~~~~~~~~~~~~~~~~~~~~~~~~~~~~~ \\ \hline \hline
    $\tilde{d}_{1}$ & $1.453 \times 10^{3}$  & $\tilde{d_{1}} \simeq (0.9910-0.0000i)\tilde{b_{L}} + (0.1289-0.0000i)\tilde{b_{R}} $\\
    $\tilde{d}_{2}$ & $1.696 \times 10^{3}$  & $\tilde{d_{2}} \simeq (0.9916-0.0000i)\tilde{b_{R}} + (-0.1286+0.0000i)\tilde{b_{L}} $\\
    $\tilde{d}_{3}$ & $1.850 \times 10^{3}$  & $\tilde{d_{3}} \simeq (0.9997+0.0189i)\tilde{s_{R}} + (0.0068+0.0001i)\tilde{s_{L}} $\\
    $\tilde{d}_{4}$ & $1.851 \times 10^{3}$  & ~~$\tilde{d_{4}} \simeq (-0.9263-0.3766i)\tilde{d_{R}} + (-0.0003-0.0001i)\tilde{d_{L}} $\\
    $\tilde{d}_{5}$ & $1.925 \times 10^{3}$  & $\tilde{d_{5}} \simeq (-0.9835-0.016i)\tilde{s_{L}} + (0.1664-0.0588i)\tilde{d_{L}} $\\
    $\tilde{d}_{6}$ & $1.926 \times 10^{3}$  & $\tilde{d_{6}} \simeq (0.8698-0.4605i)\tilde{d_{L}} + (0.1752-0.0229i)\tilde{s_{L}} $\\ \hline
    $\tilde{u}_{1}$ & $8.775 \times 10^{2}$  & $\tilde{u_{1}} \simeq (0.9604-0.0000i)\tilde{t_{R}} + (0.2749-0.0000i)\tilde{t_{L}} $ \\
    $\tilde{u}_{2}$ & $1.502 \times 10^{3}$  & $\tilde{u_{2}} \simeq (-0.9603+0.0000i)\tilde{t_{L}} + (0.2784-0.0000i)\tilde{t_{R}}$ \\
    $\tilde{u}_{3}$ & $1.858 \times 10^{3}$  & $\tilde{u_{3}} \simeq (0.9999-0.0001i)\tilde{c_{R}} + (0.0103+0.0000i)\tilde{c_{L}}$ \\
    $\tilde{u}_{4}$ & $1.858 \times 10^{3}$  & $\tilde{u_{4}} \simeq (0.2862+0.9581i)\tilde{u_{R}} + (0.0000+0.0000i)\tilde{u_{L}}$ \\
    $\tilde{u}_{5}$ & $1.924 \times 10^{3}$  & $\tilde{u_{5}} \simeq (0.9958+0.0045i)\tilde{c_{L}} + (0.0659+0.0618i)\tilde{u_{L}}$ \\
    $\tilde{u}_{6}$ & $1.924 \times 10^{3}$  & $\tilde{u_{6}} \simeq (-0.7492+0.6560i)\tilde{u_{L}} + (0.0092-0.0899i)\tilde{c_{L}}$ \\ \hline
    $\tilde{l}_{1}$ & $3.796 \times 10^{2}$  & $\tilde{l_{1}} \simeq (-0.9852+0.0000i)\tilde{\tau_{R}} + (-0.1710-0.0000i)\tilde{\tau_{L}}$\\
    $\tilde{l}_{2}$ & $7.806 \times 10^{2}$  & $\tilde{l_{2}} \simeq (-0.6766-0.7360i)\tilde{\mu_{R}} + (-0.0141-0.0154i)\tilde{\mu_{L}}$\\
    $\tilde{l}_{3}$ & $7.817 \times 10^{2}$  & $\tilde{l_{3}} \simeq (-0.6639+0.7477i)\tilde{e_{R}} + (0.0000+0.7605i)\tilde{e_{L}}$\\
    $\tilde{l}_{4}$ & $7.980 \times 10^{2}$  & $\tilde{l_{4}} \simeq (0.9852+0.0000i)\tilde{\tau_{L}} + (-0.1710-0.0000i)\tilde{\tau_{R}}$\\
    $\tilde{l}_{5}$ & $9.215 \times 10^{2}$  & $\tilde{l_{5}} \simeq (0.6681+0.7311i)\tilde{\mu_{L}} + (0.1077-0.0835i)\tilde{e_{L}}$\\
    $\tilde{l}_{6}$ & $9.219 \times 10^{2}$  & $\tilde{l_{6}} \simeq (-0.7833+0.6064i)\tilde{e_{L}} + (0.0919+0.1006i)\tilde{\mu_{L}}$\\ \hline
    $\tilde{g}$ & $1.986 \times 10^{3}$  & \\ \hline
  \end{tabular}
  \end{center}
\caption{SUSY particle masses}
 \label{SUSYpectra} 
\end{table}

The interesting prediction of MSSMRN is a small contribution to the
muon anomalous magnetic moment $g-2$, $\delta a_\mu$:
\begin{equation}
 \delta a_\mu= 3.537\times 10^{-10} \hspace{2mm}.
\end{equation}
With this contribution, the discrepancy of the theoretical value and
the experimental one becomes within 3$\sigma$, \mbox{\it i.e.} our
model satisfies a limit for $a_\mu$ at a 3 sigma level.

\subsection{Predictions for Charged LFV}
\label{subsec:lepton}

Since the slepton mixing is induced by the existence of the Dirac Yukawa couplings
via the RGE effect, we have sizable charged LFV (CLFV).

\begin{figure}[p!]
\begin{center}
\includegraphics[width=94mm]{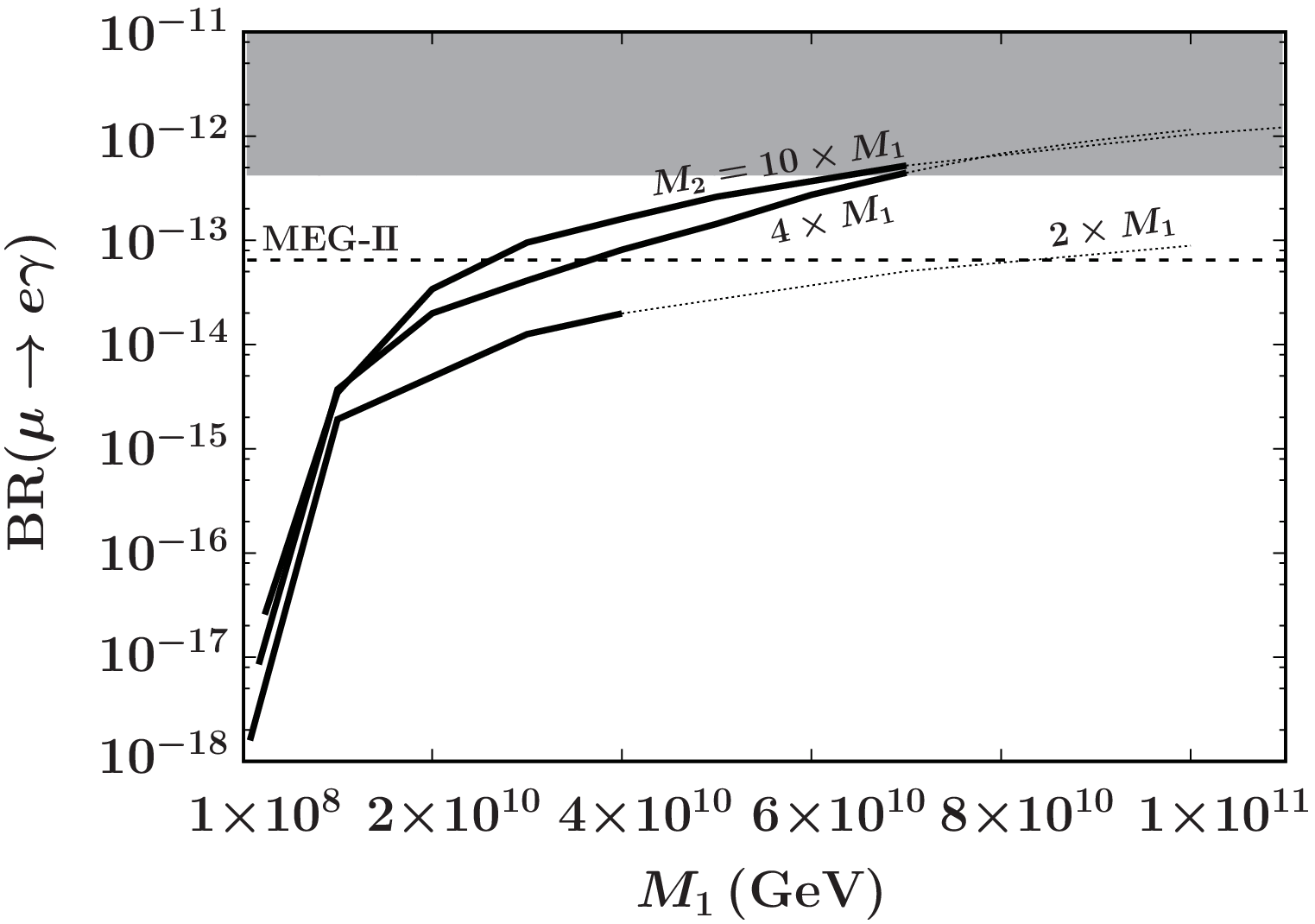}
\\[1.5mm]
\includegraphics[width=94mm]{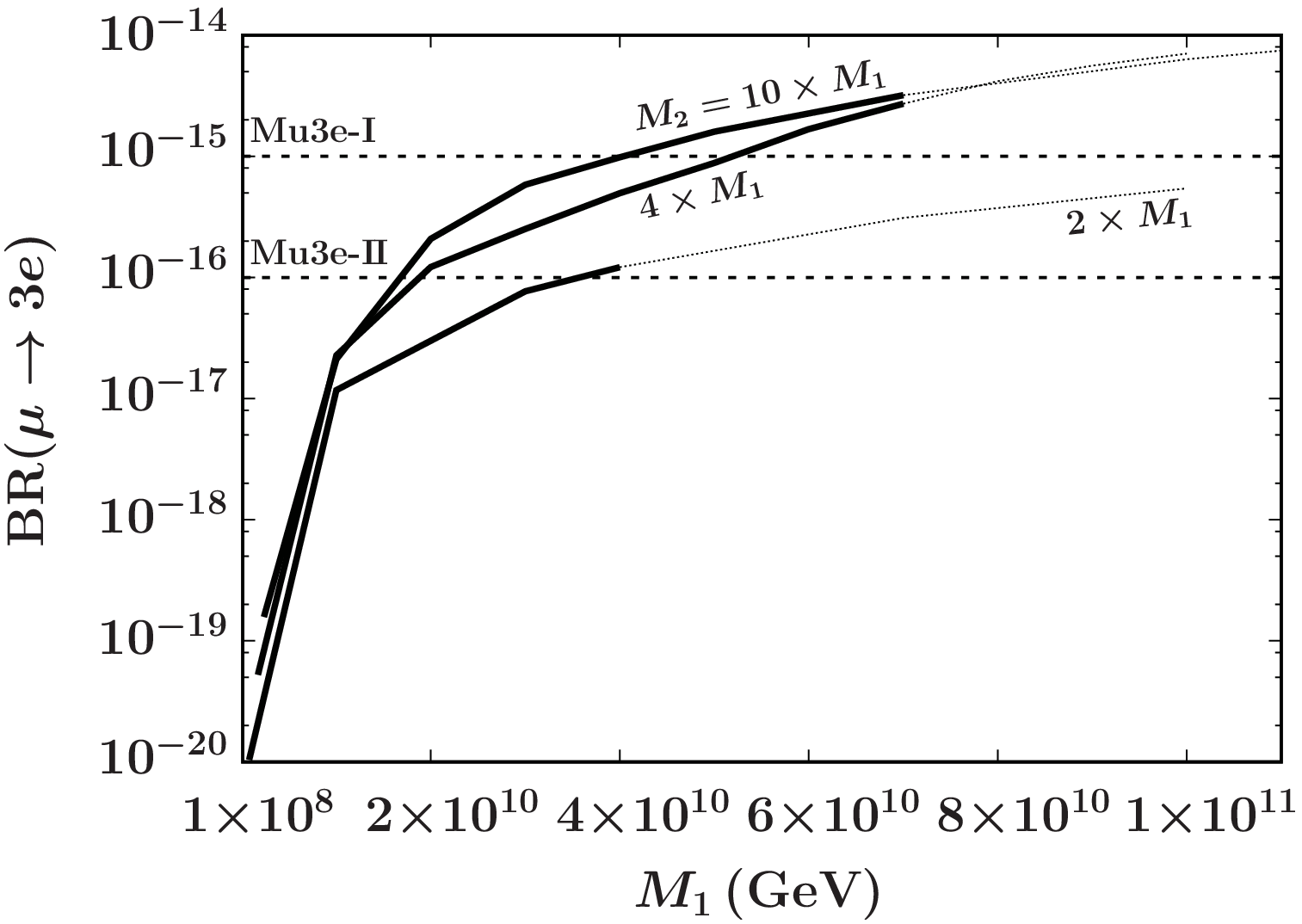}
\\[1.5mm]
\includegraphics[width=94mm]{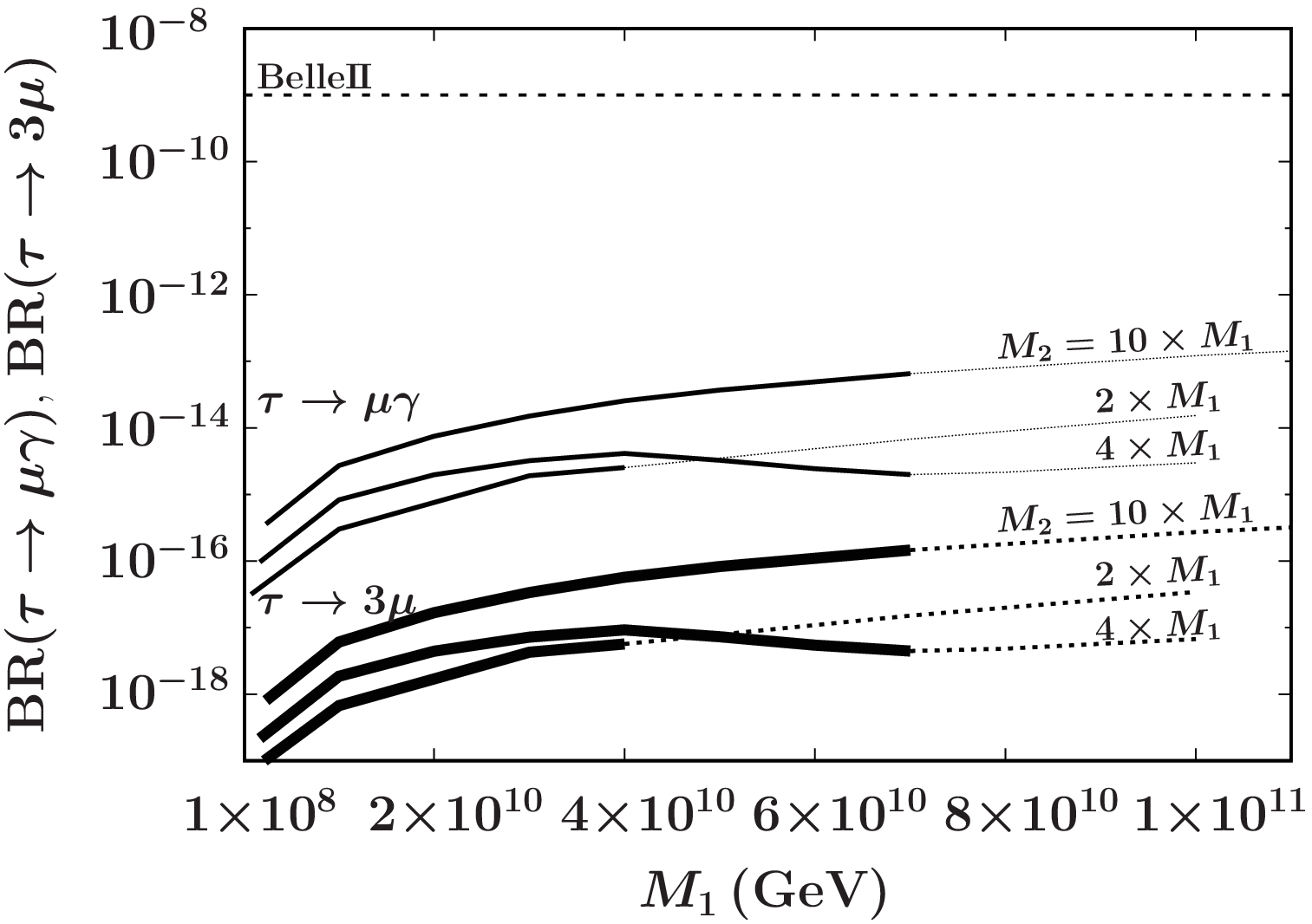}
\caption{$\text{BR}(\mu \to e \gamma)$, $\text{BR}
(\mu \to 3e)$, $\text{BR}(\tau \to \mu \gamma)$,
and $\text{BR}(\tau \to 3\mu)$ as a function of $M_{1}$
for $M_{2} = 2 \times M_{1}$, $4 \times M_{1}$, and
$10 \times M_{1}$.
The $^7$Li problem is solved with parameters in each line, 
while both the $^7$Li and $^6$Li problems are solved only for thick part.
Gray region is excluded
by MEG experiment, and the horizontal lines show future sensitivity.}
\label{Fig:M1vsBR}
\end{center}
\end{figure}

\begin{table}[htb]
\begin{center}
\vspace{2mm}
\begin{tabular}{|l|l|l|l} \hline
Process
& Bound
& Sensitivity
\\ \hline \hline
$\mu \to e \gamma$
& $4.2 \times 10^{-13}$~\cite{TheMEG:2016wtm}
& $6 \times 10^{-14}$~\cite{Baldini:2018nnn}
\\
$\mu \to 3e$
& $1.0 \times 10^{-12}$~\cite{Bellgardt:1987du}
& $1 \times 10^{-16}$~\cite{Blondel:2013ia}
\\
$\tau \to \mu \gamma$
& $4.4 \times 10^{-8}$~\cite{Aubert:2009ag}
& $1 \times 10^{-9}$~\cite{Aushev:2010bq}
\\
$\tau \to 3\mu$
& $2.1 \times 10^{-8}$~\cite{Hayasaka:2010np}
& $1 \times 10^{-9}$~\cite{Aushev:2010bq}
\\ \hline
\end{tabular}
\end{center}
\caption{Current bound and future sensitivity of branching ratio
of LFV decays.}
\label{Tab:LFV}
\end{table}

Figure~\ref{Fig:M1vsBR} shows the branching ratio of LFV decays as a function of $M_1$ 
in three different cases.  Current bound (gray region) and
future sensitivity (dashed line) are summarized in
Table~\ref{Tab:LFV}.
All of reaction rates are crudely proportional to
the second lightest Majorana neutrino mass $M_{2}$.
The dependence comes from the elements of the Dirac
neutrino Yukawa matrix $\lambda_\nu$ that have large
absolute values for a fixed active neutrino parameter
$\left|\left(\lambda_\nu\right)_{i2}\right| \propto M_{2}$.
All curves satisfy the requirement to solve the $^7$Li problem 
while the thick solid lines fullfil those for $^7$Li and $^6$Li problems. 
%

The parameter of RH neutrino is narrowed down to a small
space to solve the $^7$Li and $^6$Li problems and to generate 
successfully large lepton asymmetry. 
%
The predictions for BR$(\mu \to e \gamma)$ and BR$(\mu \to 3e)$ lie in the range
where the recent and near future experiment can probe.
Our scenario can be precisely illuminated by combining LFV
observables and unique collider signals~\cite{Biswas:2009zp,
Heisig:2011dr, Hagiwara:2012vz,
Citron:2012fg, Desai:2014uha, Heisig:2015yla, Khoze:2017ixx}.
It should be emphasized that when we consider the $^6$Li/$^7$Li
problems in the constrained MSSMRN, it is no surprise
that we have
not observed yet CLFV. 
As a matter of fact, we will observe CLFV processes in the near future.

\section{Summary and Discussion} 

We have investigated the parameter space of the constrained minimal
supersymmetric standard model with three RH Majorana neutrinos by
requiring low-energy neutrino masses and mixings. At the same time we
have applied experimental constraints such as dark matter abundance,
Li abundances, baryon asymmetry, and results from the LHC experiment,
anomalous magnetic moment and flavor observations.
We have scanned the parameter of the complex orthogonal matrix $R$ in
Eq.~(\ref{eq:eq9}) assuming a relation among the RH neutrino masses
(see Sec.~\ref{subsec:input-param}), and have found that the
allowed parameter sets really exist where all of the phenomenological
requirements are satisfied.

As shown in Sec.~\ref{sec:analysis}, it is found that the range of the
lightest RH Majorana neutrino mass $M_1$ is roughly
$10^9$ GeV $\leq M_1 \leq 10^{11}$ GeV.  The lower bound of $M_1$ is
determined to obtain sufficient amount of matter-anti-matter asymmetry
while the upper bound is determined to suppress large lepton flavor
violation.  We have also found that the degenerate mass hierarchy of
the active neutrinos are hardly realized in this region, because
rather large Yukawa couplings are necessary for the degenerate
hierarchy. The flavor mixing among sleptons are significantly canceled
through the renormalization group equation running by adjusting the complex angles. For this
reason the lightest slepton becomes long-enough-lived particle and we
thus are able to solve the $^7$Li/$^6$Li problems.

Furthermore, we have calculated the branching ratios of lepton flavor violating
decay using the allowed parameter sets.  It is found that the upper
bound of ${\rm BR}(\mu \to e \gamma)$ and ${\rm BR}(\mu \to 3e)$
are $\mathcal{O}(10^{-13})$ and $\mathcal{O}(10^{-15})$ for $M_{2}=2\times M_{1}$, 
and $\mathcal{O}(10^{-12})$ and $\mathcal{O}(10^{-14})$ for $M_{2}=4\times M_{1}$, 
respectively. The LFV decays, $\mu \to e \gamma$ and
$\mu \to 3e$, are in the reach of MEG-II and Mu3e.

\begin{acknowledgments}
  This work is supported by JSPS KAKENHI Grants No.~25105009 (J.S.),
  No.~15K17654 (T.S.), No.~16K05325 and 16K17693 (M.Y.)
\end{acknowledgments} 

\bibliographystyle{apsrev}
\bibliography{cmssmRHnu}

\begin{thebibliography}{100}
\expandafter\ifx\csname natexlab\endcsname\relax\def\natexlab#1{#1}\fi
\expandafter\ifx\csname bibnamefont\endcsname\relax
  \def\bibnamefont#1{#1}\fi
\expandafter\ifx\csname bibfnamefont\endcsname\relax
  \def\bibfnamefont#1{#1}\fi
\expandafter\ifx\csname citenamefont\endcsname\relax
  \def\citenamefont#1{#1}\fi
\expandafter\ifx\csname url\endcsname\relax
  \def\url#1{\texttt{#1}}\fi
\expandafter\ifx\csname urlprefix\endcsname\relax\def\urlprefix{URL }\fi
\providecommand{\bibinfo}[2]{#2}
\providecommand{\eprint}[2][]{\url{#2}}

\bibitem[{\citenamefont{de~Salas et~al.}(2017)\citenamefont{de~Salas, Forero,
  Ternes, Tortola, and Valle}}]{deSalas:2017kay}
\bibinfo{author}{\bibfnamefont{P.~F.} \bibnamefont{de~Salas}},
  \bibinfo{author}{\bibfnamefont{D.~V.} \bibnamefont{Forero}},
  \bibinfo{author}{\bibfnamefont{C.~A.} \bibnamefont{Ternes}},
  \bibinfo{author}{\bibfnamefont{M.}~\bibnamefont{Tortola}}, \bibnamefont{and}
  \bibinfo{author}{\bibfnamefont{J.~W.~F.} \bibnamefont{Valle}}
  (\bibinfo{year}{2017}), \eprint{1708.01186}.

\bibitem[{\citenamefont{Ade et~al.}(2016)}]{Ade:2015xua}
\bibinfo{author}{\bibfnamefont{P.~A.~R.} \bibnamefont{Ade}}
  \bibnamefont{et~al.} (\bibinfo{collaboration}{Planck}),
  \bibinfo{journal}{Astron. Astrophys.} \textbf{\bibinfo{volume}{594}},
  \bibinfo{pages}{A13} (\bibinfo{year}{2016}), \eprint{1502.01589}.

\bibitem[{\citenamefont{Aghanim et~al.}(2016)}]{Aghanim:2016yuo}
\bibinfo{author}{\bibfnamefont{N.}~\bibnamefont{Aghanim}} \bibnamefont{et~al.}
  (\bibinfo{collaboration}{Planck}), \bibinfo{journal}{Astron. Astrophys.}
  \textbf{\bibinfo{volume}{596}}, \bibinfo{pages}{A107} (\bibinfo{year}{2016}),
  \eprint{1605.02985}.

\bibitem[{\citenamefont{Minkowski}(1977)}]{Minkowski:1977sc}
\bibinfo{author}{\bibfnamefont{P.}~\bibnamefont{Minkowski}},
  \bibinfo{journal}{Phys. Lett.} \textbf{\bibinfo{volume}{67B}},
  \bibinfo{pages}{421} (\bibinfo{year}{1977}).

\bibitem[{\citenamefont{Yanagida}(1979)}]{Yanagida:1979as}
\bibinfo{author}{\bibfnamefont{T.}~\bibnamefont{Yanagida}},
  \bibinfo{journal}{Conf. Proc.} \textbf{\bibinfo{volume}{C7902131}},
  \bibinfo{pages}{95} (\bibinfo{year}{1979}).

\bibitem[{\citenamefont{Gell-Mann et~al.}(1979)\citenamefont{Gell-Mann, Ramond,
  and Slansky}}]{GellMann:1980vs}
\bibinfo{author}{\bibfnamefont{M.}~\bibnamefont{Gell-Mann}},
  \bibinfo{author}{\bibfnamefont{P.}~\bibnamefont{Ramond}}, \bibnamefont{and}
  \bibinfo{author}{\bibfnamefont{R.}~\bibnamefont{Slansky}},
  \bibinfo{journal}{Conf. Proc.} \textbf{\bibinfo{volume}{C790927}},
  \bibinfo{pages}{315} (\bibinfo{year}{1979}), \eprint{1306.4669}.

\bibitem[{\citenamefont{Glashow}(1980)}]{Glashow:1979nm}
\bibinfo{author}{\bibfnamefont{S.~L.} \bibnamefont{Glashow}},
  \bibinfo{journal}{NATO Sci. Ser. B} \textbf{\bibinfo{volume}{61}},
  \bibinfo{pages}{687} (\bibinfo{year}{1980}).

\bibitem[{\citenamefont{Mohapatra and Senjanovic}(1980)}]{Mohapatra:1979ia}
\bibinfo{author}{\bibfnamefont{R.~N.} \bibnamefont{Mohapatra}}
  \bibnamefont{and}
  \bibinfo{author}{\bibfnamefont{G.}~\bibnamefont{Senjanovic}},
  \bibinfo{journal}{Phys. Rev. Lett.} \textbf{\bibinfo{volume}{44}},
  \bibinfo{pages}{912} (\bibinfo{year}{1980}).

\bibitem[{\citenamefont{Fukugita and Yanagida}(1986)}]{Fukugita:1986hr}
\bibinfo{author}{\bibfnamefont{M.}~\bibnamefont{Fukugita}} \bibnamefont{and}
  \bibinfo{author}{\bibfnamefont{T.}~\bibnamefont{Yanagida}},
  \bibinfo{journal}{Phys. Lett.} \textbf{\bibinfo{volume}{B174}},
  \bibinfo{pages}{45} (\bibinfo{year}{1986}).

\bibitem[{\citenamefont{Kuzmin et~al.}(1985)\citenamefont{Kuzmin, Rubakov, and
  Shaposhnikov}}]{Kuzmin:1985mm}
\bibinfo{author}{\bibfnamefont{V.~A.} \bibnamefont{Kuzmin}},
  \bibinfo{author}{\bibfnamefont{V.~A.} \bibnamefont{Rubakov}},
  \bibnamefont{and} \bibinfo{author}{\bibfnamefont{M.~E.}
  \bibnamefont{Shaposhnikov}}, \bibinfo{journal}{Phys. Lett.}
  \textbf{\bibinfo{volume}{155B}}, \bibinfo{pages}{36} (\bibinfo{year}{1985}).

\bibitem[{\citenamefont{Harvey and Turner}(1990)}]{Harvey:1990qw}
\bibinfo{author}{\bibfnamefont{J.~A.} \bibnamefont{Harvey}} \bibnamefont{and}
  \bibinfo{author}{\bibfnamefont{M.~S.} \bibnamefont{Turner}},
  \bibinfo{journal}{Phys. Rev.} \textbf{\bibinfo{volume}{D42}},
  \bibinfo{pages}{3344} (\bibinfo{year}{1990}).

\bibitem[{\citenamefont{Buchmuller et~al.}(2002)\citenamefont{Buchmuller,
  Di~Bari, and Plumacher}}]{Buchmuller:2002rq}
\bibinfo{author}{\bibfnamefont{W.}~\bibnamefont{Buchmuller}},
  \bibinfo{author}{\bibfnamefont{P.}~\bibnamefont{Di~Bari}}, \bibnamefont{and}
  \bibinfo{author}{\bibfnamefont{M.}~\bibnamefont{Plumacher}},
  \bibinfo{journal}{Nucl. Phys.} \textbf{\bibinfo{volume}{B643}},
  \bibinfo{pages}{367} (\bibinfo{year}{2002}), \bibinfo{note}{[Erratum: Nucl.
  Phys.B793,362(2008)]}, \eprint{hep-ph/0205349}.

\bibitem[{\citenamefont{Ellis and Raidal}(2002)}]{Ellis:2002xg}
\bibinfo{author}{\bibfnamefont{J.~R.} \bibnamefont{Ellis}} \bibnamefont{and}
  \bibinfo{author}{\bibfnamefont{M.}~\bibnamefont{Raidal}},
  \bibinfo{journal}{Nucl. Phys.} \textbf{\bibinfo{volume}{B643}},
  \bibinfo{pages}{229} (\bibinfo{year}{2002}), \eprint{hep-ph/0206174}.

\bibitem[{\citenamefont{Bando et~al.}(2004)\citenamefont{Bando, Kaneko, Obara,
  and Tanimoto}}]{Bando:2004hi}
\bibinfo{author}{\bibfnamefont{M.}~\bibnamefont{Bando}},
  \bibinfo{author}{\bibfnamefont{S.}~\bibnamefont{Kaneko}},
  \bibinfo{author}{\bibfnamefont{M.}~\bibnamefont{Obara}}, \bibnamefont{and}
  \bibinfo{author}{\bibfnamefont{M.}~\bibnamefont{Tanimoto}},
  \bibinfo{journal}{Prog. Theor. Phys.} \textbf{\bibinfo{volume}{112}},
  \bibinfo{pages}{533} (\bibinfo{year}{2004}), \eprint{hep-ph/0405071}.

\bibitem[{\citenamefont{Chang et~al.}(2004)\citenamefont{Chang, Kang, and
  Siyeon}}]{Chang:2004wy}
\bibinfo{author}{\bibfnamefont{S.}~\bibnamefont{Chang}},
  \bibinfo{author}{\bibfnamefont{S.~K.} \bibnamefont{Kang}}, \bibnamefont{and}
  \bibinfo{author}{\bibfnamefont{K.}~\bibnamefont{Siyeon}},
  \bibinfo{journal}{Phys. Lett.} \textbf{\bibinfo{volume}{B597}},
  \bibinfo{pages}{78} (\bibinfo{year}{2004}), \eprint{hep-ph/0404187}.

\bibitem[{\citenamefont{Petcov et~al.}(2006{\natexlab{a}})\citenamefont{Petcov,
  Rodejohann, Shindou, and Takanishi}}]{Petcov:2005jh}
\bibinfo{author}{\bibfnamefont{S.~T.} \bibnamefont{Petcov}},
  \bibinfo{author}{\bibfnamefont{W.}~\bibnamefont{Rodejohann}},
  \bibinfo{author}{\bibfnamefont{T.}~\bibnamefont{Shindou}}, \bibnamefont{and}
  \bibinfo{author}{\bibfnamefont{Y.}~\bibnamefont{Takanishi}},
  \bibinfo{journal}{Nucl. Phys.} \textbf{\bibinfo{volume}{B739}},
  \bibinfo{pages}{208} (\bibinfo{year}{2006}{\natexlab{a}}),
  \eprint{hep-ph/0510404}.

\bibitem[{\citenamefont{Guo et~al.}(2007)\citenamefont{Guo, Xing, and
  Zhou}}]{Guo:2006qa}
\bibinfo{author}{\bibfnamefont{W.-l.} \bibnamefont{Guo}},
  \bibinfo{author}{\bibfnamefont{Z.-z.} \bibnamefont{Xing}}, \bibnamefont{and}
  \bibinfo{author}{\bibfnamefont{S.}~\bibnamefont{Zhou}},
  \bibinfo{journal}{Int. J. Mod. Phys.} \textbf{\bibinfo{volume}{E16}},
  \bibinfo{pages}{1} (\bibinfo{year}{2007}), \eprint{hep-ph/0612033}.

\bibitem[{\citenamefont{Pascoli et~al.}(2007)\citenamefont{Pascoli, Petcov, and
  Riotto}}]{Pascoli:2006ci}
\bibinfo{author}{\bibfnamefont{S.}~\bibnamefont{Pascoli}},
  \bibinfo{author}{\bibfnamefont{S.~T.} \bibnamefont{Petcov}},
  \bibnamefont{and} \bibinfo{author}{\bibfnamefont{A.}~\bibnamefont{Riotto}},
  \bibinfo{journal}{Nucl. Phys.} \textbf{\bibinfo{volume}{B774}},
  \bibinfo{pages}{1} (\bibinfo{year}{2007}), \eprint{hep-ph/0611338}.

\bibitem[{\citenamefont{Spergel et~al.}(2007)}]{Spergel:2006hy}
\bibinfo{author}{\bibfnamefont{D.~N.} \bibnamefont{Spergel}}
  \bibnamefont{et~al.} (\bibinfo{collaboration}{WMAP}),
  \bibinfo{journal}{Astrophys. J. Suppl.} \textbf{\bibinfo{volume}{170}},
  \bibinfo{pages}{377} (\bibinfo{year}{2007}), \eprint{astro-ph/0603449}.

\bibitem[{\citenamefont{Griest and Seckel}(1991)}]{Griest:1990kh}
\bibinfo{author}{\bibfnamefont{K.}~\bibnamefont{Griest}} \bibnamefont{and}
  \bibinfo{author}{\bibfnamefont{D.}~\bibnamefont{Seckel}},
  \bibinfo{journal}{Phys. Rev.} \textbf{\bibinfo{volume}{D43}},
  \bibinfo{pages}{3191} (\bibinfo{year}{1991}).

\bibitem[{\citenamefont{Profumo et~al.}(2005)\citenamefont{Profumo, Sigurdson,
  Ullio, and Kamionkowski}}]{Profumo:2004qt}
\bibinfo{author}{\bibfnamefont{S.}~\bibnamefont{Profumo}},
  \bibinfo{author}{\bibfnamefont{K.}~\bibnamefont{Sigurdson}},
  \bibinfo{author}{\bibfnamefont{P.}~\bibnamefont{Ullio}}, \bibnamefont{and}
  \bibinfo{author}{\bibfnamefont{M.}~\bibnamefont{Kamionkowski}},
  \bibinfo{journal}{Phys. Rev.} \textbf{\bibinfo{volume}{D71}},
  \bibinfo{pages}{023518} (\bibinfo{year}{2005}), \eprint{astro-ph/0410714}.

\bibitem[{\citenamefont{Gladyshev et~al.}(2005)\citenamefont{Gladyshev,
  Kazakov, and Paucar}}]{Gladyshev:2005mn}
\bibinfo{author}{\bibfnamefont{A.~V.} \bibnamefont{Gladyshev}},
  \bibinfo{author}{\bibfnamefont{D.~I.} \bibnamefont{Kazakov}},
  \bibnamefont{and} \bibinfo{author}{\bibfnamefont{M.~G.}
  \bibnamefont{Paucar}}, \bibinfo{journal}{Mod. Phys. Lett.}
  \textbf{\bibinfo{volume}{A20}}, \bibinfo{pages}{3085} (\bibinfo{year}{2005}),
  \eprint{hep-ph/0509168}.

\bibitem[{\citenamefont{Jittoh et~al.}(2006)\citenamefont{Jittoh, Sato,
  Shimomura, and Yamanaka}}]{Jittoh:2005pq}
\bibinfo{author}{\bibfnamefont{T.}~\bibnamefont{Jittoh}},
  \bibinfo{author}{\bibfnamefont{J.}~\bibnamefont{Sato}},
  \bibinfo{author}{\bibfnamefont{T.}~\bibnamefont{Shimomura}},
  \bibnamefont{and} \bibinfo{author}{\bibfnamefont{M.}~\bibnamefont{Yamanaka}},
  \bibinfo{journal}{Phys. Rev.} \textbf{\bibinfo{volume}{D73}},
  \bibinfo{pages}{055009} (\bibinfo{year}{2006}), \bibinfo{note}{[Erratum:
  Phys. Rev.D87,no.1,019901(2013)]}, \eprint{hep-ph/0512197}.

\bibitem[{\citenamefont{Cyburt et~al.}(2008)\citenamefont{Cyburt, Fields, and
  Olive}}]{Cyburt:2008kw}
\bibinfo{author}{\bibfnamefont{R.~H.} \bibnamefont{Cyburt}},
  \bibinfo{author}{\bibfnamefont{B.~D.} \bibnamefont{Fields}},
  \bibnamefont{and} \bibinfo{author}{\bibfnamefont{K.~A.} \bibnamefont{Olive}},
  \bibinfo{journal}{JCAP} \textbf{\bibinfo{volume}{0811}}, \bibinfo{pages}{012}
  (\bibinfo{year}{2008}), \eprint{0808.2818}.

\bibitem[{\citenamefont{Sbordone et~al.}(2010)}]{Sbordone:2010zi}
\bibinfo{author}{\bibfnamefont{L.}~\bibnamefont{Sbordone}}
  \bibnamefont{et~al.}, \bibinfo{journal}{Astron. Astrophys.}
  \textbf{\bibinfo{volume}{522}}, \bibinfo{pages}{A26} (\bibinfo{year}{2010}),
  \eprint{1003.4510}.

\bibitem[{\citenamefont{Coc et~al.}(2012)\citenamefont{Coc, Goriely, Xu,
  Saimpert, and Vangioni}}]{Coc:2011az}
\bibinfo{author}{\bibfnamefont{A.}~\bibnamefont{Coc}},
  \bibinfo{author}{\bibfnamefont{S.}~\bibnamefont{Goriely}},
  \bibinfo{author}{\bibfnamefont{Y.}~\bibnamefont{Xu}},
  \bibinfo{author}{\bibfnamefont{M.}~\bibnamefont{Saimpert}}, \bibnamefont{and}
  \bibinfo{author}{\bibfnamefont{E.}~\bibnamefont{Vangioni}},
  \bibinfo{journal}{Astrophys. J.} \textbf{\bibinfo{volume}{744}},
  \bibinfo{pages}{158} (\bibinfo{year}{2012}), \eprint{1107.1117}.

\bibitem[{\citenamefont{Asplund et~al.}(2006)\citenamefont{Asplund, Lambert,
  Nissen, Primas, and Smith}}]{Asplund:2005yt}
\bibinfo{author}{\bibfnamefont{M.}~\bibnamefont{Asplund}},
  \bibinfo{author}{\bibfnamefont{D.~L.} \bibnamefont{Lambert}},
  \bibinfo{author}{\bibfnamefont{P.~E.} \bibnamefont{Nissen}},
  \bibinfo{author}{\bibfnamefont{F.}~\bibnamefont{Primas}}, \bibnamefont{and}
  \bibinfo{author}{\bibfnamefont{V.~V.} \bibnamefont{Smith}},
  \bibinfo{journal}{Astrophys. J.} \textbf{\bibinfo{volume}{644}},
  \bibinfo{pages}{229} (\bibinfo{year}{2006}), \eprint{astro-ph/0510636}.

\bibitem[{\citenamefont{Kawabata et~al.}(2017)}]{Kawabata:2017zpa}
\bibinfo{author}{\bibfnamefont{T.}~\bibnamefont{Kawabata}}
  \bibnamefont{et~al.}, \bibinfo{journal}{Phys. Rev. Lett.}
  \textbf{\bibinfo{volume}{118}}, \bibinfo{pages}{052701}
  (\bibinfo{year}{2017}).

\bibitem[{\citenamefont{Jittoh et~al.}(2007)\citenamefont{Jittoh, Kohri, Koike,
  Sato, Shimomura, and Yamanaka}}]{Jittoh:2007fr}
\bibinfo{author}{\bibfnamefont{T.}~\bibnamefont{Jittoh}},
  \bibinfo{author}{\bibfnamefont{K.}~\bibnamefont{Kohri}},
  \bibinfo{author}{\bibfnamefont{M.}~\bibnamefont{Koike}},
  \bibinfo{author}{\bibfnamefont{J.}~\bibnamefont{Sato}},
  \bibinfo{author}{\bibfnamefont{T.}~\bibnamefont{Shimomura}},
  \bibnamefont{and} \bibinfo{author}{\bibfnamefont{M.}~\bibnamefont{Yamanaka}},
  \bibinfo{journal}{Phys. Rev.} \textbf{\bibinfo{volume}{D76}},
  \bibinfo{pages}{125023} (\bibinfo{year}{2007}), \eprint{0704.2914}.

\bibitem[{\citenamefont{Jittoh et~al.}(2008)\citenamefont{Jittoh, Kohri, Koike,
  Sato, Shimomura, and Yamanaka}}]{Jittoh:2008eq}
\bibinfo{author}{\bibfnamefont{T.}~\bibnamefont{Jittoh}},
  \bibinfo{author}{\bibfnamefont{K.}~\bibnamefont{Kohri}},
  \bibinfo{author}{\bibfnamefont{M.}~\bibnamefont{Koike}},
  \bibinfo{author}{\bibfnamefont{J.}~\bibnamefont{Sato}},
  \bibinfo{author}{\bibfnamefont{T.}~\bibnamefont{Shimomura}},
  \bibnamefont{and} \bibinfo{author}{\bibfnamefont{M.}~\bibnamefont{Yamanaka}},
  \bibinfo{journal}{Phys. Rev.} \textbf{\bibinfo{volume}{D78}},
  \bibinfo{pages}{055007} (\bibinfo{year}{2008}), \eprint{0805.3389}.

\bibitem[{\citenamefont{Pospelov}(2007)}]{Pospelov:2006sc}
\bibinfo{author}{\bibfnamefont{M.}~\bibnamefont{Pospelov}},
  \bibinfo{journal}{Phys. Rev. Lett.} \textbf{\bibinfo{volume}{98}},
  \bibinfo{pages}{231301} (\bibinfo{year}{2007}), \eprint{hep-ph/0605215}.

\bibitem[{\citenamefont{Konishi et~al.}(2014)\citenamefont{Konishi, Ohta, Sato,
  Shimomura, Sugai, and Yamanaka}}]{Konishi:2013gda}
\bibinfo{author}{\bibfnamefont{Y.}~\bibnamefont{Konishi}},
  \bibinfo{author}{\bibfnamefont{S.}~\bibnamefont{Ohta}},
  \bibinfo{author}{\bibfnamefont{J.}~\bibnamefont{Sato}},
  \bibinfo{author}{\bibfnamefont{T.}~\bibnamefont{Shimomura}},
  \bibinfo{author}{\bibfnamefont{K.}~\bibnamefont{Sugai}}, \bibnamefont{and}
  \bibinfo{author}{\bibfnamefont{M.}~\bibnamefont{Yamanaka}},
  \bibinfo{journal}{Phys. Rev.} \textbf{\bibinfo{volume}{D89}},
  \bibinfo{pages}{075006} (\bibinfo{year}{2014}), \eprint{1309.2067}.

\bibitem[{\citenamefont{Borzumati and Masiero}(1986)}]{Borzumati:1986qx}
\bibinfo{author}{\bibfnamefont{F.}~\bibnamefont{Borzumati}} \bibnamefont{and}
  \bibinfo{author}{\bibfnamefont{A.}~\bibnamefont{Masiero}},
  \bibinfo{journal}{Phys. Rev. Lett.} \textbf{\bibinfo{volume}{57}},
  \bibinfo{pages}{961} (\bibinfo{year}{1986}).

\bibitem[{\citenamefont{Hisano et~al.}(1996)\citenamefont{Hisano, Moroi, Tobe,
  and Yamaguchi}}]{Hisano:1995cp}
\bibinfo{author}{\bibfnamefont{J.}~\bibnamefont{Hisano}},
  \bibinfo{author}{\bibfnamefont{T.}~\bibnamefont{Moroi}},
  \bibinfo{author}{\bibfnamefont{K.}~\bibnamefont{Tobe}}, \bibnamefont{and}
  \bibinfo{author}{\bibfnamefont{M.}~\bibnamefont{Yamaguchi}},
  \bibinfo{journal}{Phys. Rev.} \textbf{\bibinfo{volume}{D53}},
  \bibinfo{pages}{2442} (\bibinfo{year}{1996}), \eprint{hep-ph/9510309}.

\bibitem[{\citenamefont{Casas and Ibarra}(2001)}]{Casas:2001sr}
\bibinfo{author}{\bibfnamefont{J.~A.} \bibnamefont{Casas}} \bibnamefont{and}
  \bibinfo{author}{\bibfnamefont{A.}~\bibnamefont{Ibarra}},
  \bibinfo{journal}{Nucl. Phys.} \textbf{\bibinfo{volume}{B618}},
  \bibinfo{pages}{171} (\bibinfo{year}{2001}), \eprint{hep-ph/0103065}.

\bibitem[{\citenamefont{Ellis et~al.}(2002{\natexlab{a}})\citenamefont{Ellis,
  Hisano, Lola, and Raidal}}]{Ellis:2001xt}
\bibinfo{author}{\bibfnamefont{J.~R.} \bibnamefont{Ellis}},
  \bibinfo{author}{\bibfnamefont{J.}~\bibnamefont{Hisano}},
  \bibinfo{author}{\bibfnamefont{S.}~\bibnamefont{Lola}}, \bibnamefont{and}
  \bibinfo{author}{\bibfnamefont{M.}~\bibnamefont{Raidal}},
  \bibinfo{journal}{Nucl. Phys.} \textbf{\bibinfo{volume}{B621}},
  \bibinfo{pages}{208} (\bibinfo{year}{2002}{\natexlab{a}}),
  \eprint{hep-ph/0109125}.

\bibitem[{\citenamefont{Ellis et~al.}(2002{\natexlab{b}})\citenamefont{Ellis,
  Hisano, Raidal, and Shimizu}}]{Ellis:2002fe}
\bibinfo{author}{\bibfnamefont{J.~R.} \bibnamefont{Ellis}},
  \bibinfo{author}{\bibfnamefont{J.}~\bibnamefont{Hisano}},
  \bibinfo{author}{\bibfnamefont{M.}~\bibnamefont{Raidal}}, \bibnamefont{and}
  \bibinfo{author}{\bibfnamefont{Y.}~\bibnamefont{Shimizu}},
  \bibinfo{journal}{Phys. Rev.} \textbf{\bibinfo{volume}{D66}},
  \bibinfo{pages}{115013} (\bibinfo{year}{2002}{\natexlab{b}}),
  \eprint{hep-ph/0206110}.

\bibitem[{\citenamefont{Lavignac et~al.}(2001)\citenamefont{Lavignac, Masina,
  and Savoy}}]{Lavignac:2001vp}
\bibinfo{author}{\bibfnamefont{S.}~\bibnamefont{Lavignac}},
  \bibinfo{author}{\bibfnamefont{I.}~\bibnamefont{Masina}}, \bibnamefont{and}
  \bibinfo{author}{\bibfnamefont{C.~A.} \bibnamefont{Savoy}},
  \bibinfo{journal}{Phys. Lett.} \textbf{\bibinfo{volume}{B520}},
  \bibinfo{pages}{269} (\bibinfo{year}{2001}), \eprint{hep-ph/0106245}.

\bibitem[{\citenamefont{Kageyama et~al.}(2002)\citenamefont{Kageyama, Kaneko,
  Shimoyama, and Tanimoto}}]{Kageyama:2001tn}
\bibinfo{author}{\bibfnamefont{A.}~\bibnamefont{Kageyama}},
  \bibinfo{author}{\bibfnamefont{S.}~\bibnamefont{Kaneko}},
  \bibinfo{author}{\bibfnamefont{N.}~\bibnamefont{Shimoyama}},
  \bibnamefont{and} \bibinfo{author}{\bibfnamefont{M.}~\bibnamefont{Tanimoto}},
  \bibinfo{journal}{Phys. Rev.} \textbf{\bibinfo{volume}{D65}},
  \bibinfo{pages}{096010} (\bibinfo{year}{2002}), \eprint{hep-ph/0112359}.

\bibitem[{\citenamefont{Deppisch et~al.}(2003)\citenamefont{Deppisch, Pas,
  Redelbach, Ruckl, and Shimizu}}]{Deppisch:2002vz}
\bibinfo{author}{\bibfnamefont{F.}~\bibnamefont{Deppisch}},
  \bibinfo{author}{\bibfnamefont{H.}~\bibnamefont{Pas}},
  \bibinfo{author}{\bibfnamefont{A.}~\bibnamefont{Redelbach}},
  \bibinfo{author}{\bibfnamefont{R.}~\bibnamefont{Ruckl}}, \bibnamefont{and}
  \bibinfo{author}{\bibfnamefont{Y.}~\bibnamefont{Shimizu}},
  \bibinfo{journal}{Eur. Phys. J.} \textbf{\bibinfo{volume}{C28}},
  \bibinfo{pages}{365} (\bibinfo{year}{2003}), \eprint{hep-ph/0206122}.

\bibitem[{\citenamefont{Blazek and King}(2003)}]{Blazek:2002wq}
\bibinfo{author}{\bibfnamefont{T.}~\bibnamefont{Blazek}} \bibnamefont{and}
  \bibinfo{author}{\bibfnamefont{S.~F.} \bibnamefont{King}},
  \bibinfo{journal}{Nucl. Phys.} \textbf{\bibinfo{volume}{B662}},
  \bibinfo{pages}{359} (\bibinfo{year}{2003}), \eprint{hep-ph/0211368}.

\bibitem[{\citenamefont{Petcov et~al.}(2004)\citenamefont{Petcov, Profumo,
  Takanishi, and Yaguna}}]{Petcov:2003zb}
\bibinfo{author}{\bibfnamefont{S.~T.} \bibnamefont{Petcov}},
  \bibinfo{author}{\bibfnamefont{S.}~\bibnamefont{Profumo}},
  \bibinfo{author}{\bibfnamefont{Y.}~\bibnamefont{Takanishi}},
  \bibnamefont{and} \bibinfo{author}{\bibfnamefont{C.~E.}
  \bibnamefont{Yaguna}}, \bibinfo{journal}{Nucl. Phys.}
  \textbf{\bibinfo{volume}{B676}}, \bibinfo{pages}{453} (\bibinfo{year}{2004}),
  \eprint{hep-ph/0306195}.

\bibitem[{\citenamefont{Dutta and Mohapatra}(2003)}]{Dutta:2003ps}
\bibinfo{author}{\bibfnamefont{B.}~\bibnamefont{Dutta}} \bibnamefont{and}
  \bibinfo{author}{\bibfnamefont{R.~N.} \bibnamefont{Mohapatra}},
  \bibinfo{journal}{Phys. Rev.} \textbf{\bibinfo{volume}{D68}},
  \bibinfo{pages}{056006} (\bibinfo{year}{2003}), \eprint{hep-ph/0305059}.

\bibitem[{\citenamefont{Illana and Masip}(2004)}]{Illana:2003pj}
\bibinfo{author}{\bibfnamefont{J.~I.} \bibnamefont{Illana}} \bibnamefont{and}
  \bibinfo{author}{\bibfnamefont{M.}~\bibnamefont{Masip}},
  \bibinfo{journal}{Eur. Phys. J.} \textbf{\bibinfo{volume}{C35}},
  \bibinfo{pages}{365} (\bibinfo{year}{2004}), \eprint{hep-ph/0307393}.

\bibitem[{\citenamefont{Babu et~al.}(2005)\citenamefont{Babu, Pati, and
  Rastogi}}]{Babu:2005yr}
\bibinfo{author}{\bibfnamefont{K.~S.} \bibnamefont{Babu}},
  \bibinfo{author}{\bibfnamefont{J.~C.} \bibnamefont{Pati}}, \bibnamefont{and}
  \bibinfo{author}{\bibfnamefont{P.}~\bibnamefont{Rastogi}},
  \bibinfo{journal}{Phys. Lett.} \textbf{\bibinfo{volume}{B621}},
  \bibinfo{pages}{160} (\bibinfo{year}{2005}), \eprint{hep-ph/0502152}.

\bibitem[{\citenamefont{Petcov et~al.}(2006{\natexlab{b}})\citenamefont{Petcov,
  Shindou, and Takanishi}}]{Petcov:2005yh}
\bibinfo{author}{\bibfnamefont{S.~T.} \bibnamefont{Petcov}},
  \bibinfo{author}{\bibfnamefont{T.}~\bibnamefont{Shindou}}, \bibnamefont{and}
  \bibinfo{author}{\bibfnamefont{Y.}~\bibnamefont{Takanishi}},
  \bibinfo{journal}{Nucl. Phys.} \textbf{\bibinfo{volume}{B738}},
  \bibinfo{pages}{219} (\bibinfo{year}{2006}{\natexlab{b}}),
  \eprint{hep-ph/0508243}.

\bibitem[{\citenamefont{Calibbi and Signorelli}(2018)}]{Calibbi:2017uvl}
\bibinfo{author}{\bibfnamefont{L.}~\bibnamefont{Calibbi}} \bibnamefont{and}
  \bibinfo{author}{\bibfnamefont{G.}~\bibnamefont{Signorelli}},
  \bibinfo{journal}{Riv. Nuovo Cim.} \textbf{\bibinfo{volume}{41}},
  \bibinfo{pages}{1} (\bibinfo{year}{2018}), \eprint{1709.00294}.

\bibitem[{\citenamefont{Patrignani et~al.}(2016)}]{Patrignani:2016xqp}
\bibinfo{author}{\bibfnamefont{C.}~\bibnamefont{Patrignani}}
  \bibnamefont{et~al.} (\bibinfo{collaboration}{Particle Data Group}),
  \bibinfo{journal}{Chin. Phys.} \textbf{\bibinfo{volume}{C40}},
  \bibinfo{pages}{100001} (\bibinfo{year}{2016}).

\bibitem[{\citenamefont{Kaneko et~al.}(2013)\citenamefont{Kaneko, Sato,
  Shimomura, Vives, and Yamanaka}}]{Kaneko:2008re}
\bibinfo{author}{\bibfnamefont{S.}~\bibnamefont{Kaneko}},
  \bibinfo{author}{\bibfnamefont{J.}~\bibnamefont{Sato}},
  \bibinfo{author}{\bibfnamefont{T.}~\bibnamefont{Shimomura}},
  \bibinfo{author}{\bibfnamefont{O.}~\bibnamefont{Vives}}, \bibnamefont{and}
  \bibinfo{author}{\bibfnamefont{M.}~\bibnamefont{Yamanaka}},
  \bibinfo{journal}{Phys. Rev.} \textbf{\bibinfo{volume}{D87}},
  \bibinfo{pages}{039904} (\bibinfo{year}{2013}), \bibinfo{note}{[Phys.
  Rev.D78,no.11,116013(2008)]}, \eprint{0811.0703}.

\bibitem[{\citenamefont{Kaneko et~al.}(2011)\citenamefont{Kaneko, Saito, Sato,
  Shimomura, Vives, and Yamanaka}}]{Kaneko:2011qi}
\bibinfo{author}{\bibfnamefont{S.}~\bibnamefont{Kaneko}},
  \bibinfo{author}{\bibfnamefont{H.}~\bibnamefont{Saito}},
  \bibinfo{author}{\bibfnamefont{J.}~\bibnamefont{Sato}},
  \bibinfo{author}{\bibfnamefont{T.}~\bibnamefont{Shimomura}},
  \bibinfo{author}{\bibfnamefont{O.}~\bibnamefont{Vives}}, \bibnamefont{and}
  \bibinfo{author}{\bibfnamefont{M.}~\bibnamefont{Yamanaka}},
  \bibinfo{journal}{Phys. Rev.} \textbf{\bibinfo{volume}{D83}},
  \bibinfo{pages}{115005} (\bibinfo{year}{2011}), \eprint{1102.1794}.

\bibitem[{\citenamefont{Edsjo and Gondolo}(1997)}]{Edsjo:1997bg}
\bibinfo{author}{\bibfnamefont{J.}~\bibnamefont{Edsjo}} \bibnamefont{and}
  \bibinfo{author}{\bibfnamefont{P.}~\bibnamefont{Gondolo}},
  \bibinfo{journal}{Phys. Rev.} \textbf{\bibinfo{volume}{D56}},
  \bibinfo{pages}{1879} (\bibinfo{year}{1997}), \eprint{hep-ph/9704361}.

\bibitem[{\citenamefont{Nihei et~al.}(2002)\citenamefont{Nihei, Roszkowski, and
  Ruiz~de Austri}}]{Nihei:2002sc}
\bibinfo{author}{\bibfnamefont{T.}~\bibnamefont{Nihei}},
  \bibinfo{author}{\bibfnamefont{L.}~\bibnamefont{Roszkowski}},
  \bibnamefont{and} \bibinfo{author}{\bibfnamefont{R.}~\bibnamefont{Ruiz~de
  Austri}}, \bibinfo{journal}{JHEP} \textbf{\bibinfo{volume}{07}},
  \bibinfo{pages}{024} (\bibinfo{year}{2002}), \eprint{hep-ph/0206266}.

\bibitem[{\citenamefont{Kohri et~al.}(2012)\citenamefont{Kohri, Ohta, Sato,
  Shimomura, and Yamanaka}}]{Kohri:2012gc}
\bibinfo{author}{\bibfnamefont{K.}~\bibnamefont{Kohri}},
  \bibinfo{author}{\bibfnamefont{S.}~\bibnamefont{Ohta}},
  \bibinfo{author}{\bibfnamefont{J.}~\bibnamefont{Sato}},
  \bibinfo{author}{\bibfnamefont{T.}~\bibnamefont{Shimomura}},
  \bibnamefont{and} \bibinfo{author}{\bibfnamefont{M.}~\bibnamefont{Yamanaka}},
  \bibinfo{journal}{Phys. Rev.} \textbf{\bibinfo{volume}{D86}},
  \bibinfo{pages}{095024} (\bibinfo{year}{2012}), \eprint{1208.5533}.

\bibitem[{\citenamefont{Jittoh et~al.}(2010)\citenamefont{Jittoh, Kohri, Koike,
  Sato, Shimomura, and Yamanaka}}]{Jittoh:2010wh}
\bibinfo{author}{\bibfnamefont{T.}~\bibnamefont{Jittoh}},
  \bibinfo{author}{\bibfnamefont{K.}~\bibnamefont{Kohri}},
  \bibinfo{author}{\bibfnamefont{M.}~\bibnamefont{Koike}},
  \bibinfo{author}{\bibfnamefont{J.}~\bibnamefont{Sato}},
  \bibinfo{author}{\bibfnamefont{T.}~\bibnamefont{Shimomura}},
  \bibnamefont{and} \bibinfo{author}{\bibfnamefont{M.}~\bibnamefont{Yamanaka}},
  \bibinfo{journal}{Phys. Rev.} \textbf{\bibinfo{volume}{D82}},
  \bibinfo{pages}{115030} (\bibinfo{year}{2010}), \eprint{1001.1217}.

\bibitem[{\citenamefont{Jittoh et~al.}(2011)\citenamefont{Jittoh, Kohri, Koike,
  Sato, Sugai, Yamanaka, and Yazaki}}]{Jittoh:2011ni}
\bibinfo{author}{\bibfnamefont{T.}~\bibnamefont{Jittoh}},
  \bibinfo{author}{\bibfnamefont{K.}~\bibnamefont{Kohri}},
  \bibinfo{author}{\bibfnamefont{M.}~\bibnamefont{Koike}},
  \bibinfo{author}{\bibfnamefont{J.}~\bibnamefont{Sato}},
  \bibinfo{author}{\bibfnamefont{K.}~\bibnamefont{Sugai}},
  \bibinfo{author}{\bibfnamefont{M.}~\bibnamefont{Yamanaka}}, \bibnamefont{and}
  \bibinfo{author}{\bibfnamefont{K.}~\bibnamefont{Yazaki}},
  \bibinfo{journal}{Phys. Rev.} \textbf{\bibinfo{volume}{D84}},
  \bibinfo{pages}{035008} (\bibinfo{year}{2011}), \eprint{1105.1431}.

\bibitem[{\citenamefont{Kohri et~al.}(2014)\citenamefont{Kohri, Koike, Konishi,
  Ohta, Sato, Shimomura, Sugai, and Yamanaka}}]{Kohri:2014jfa}
\bibinfo{author}{\bibfnamefont{K.}~\bibnamefont{Kohri}},
  \bibinfo{author}{\bibfnamefont{M.}~\bibnamefont{Koike}},
  \bibinfo{author}{\bibfnamefont{Y.}~\bibnamefont{Konishi}},
  \bibinfo{author}{\bibfnamefont{S.}~\bibnamefont{Ohta}},
  \bibinfo{author}{\bibfnamefont{J.}~\bibnamefont{Sato}},
  \bibinfo{author}{\bibfnamefont{T.}~\bibnamefont{Shimomura}},
  \bibinfo{author}{\bibfnamefont{K.}~\bibnamefont{Sugai}}, \bibnamefont{and}
  \bibinfo{author}{\bibfnamefont{M.}~\bibnamefont{Yamanaka}},
  \bibinfo{journal}{Phys. Rev.} \textbf{\bibinfo{volume}{D90}},
  \bibinfo{pages}{035003} (\bibinfo{year}{2014}), \eprint{1403.1561}.

\bibitem[{\citenamefont{Kusakabe et~al.}(2010)\citenamefont{Kusakabe, Kajino,
  Yoshida, and Mathews}}]{Kusakabe:2010cb}
\bibinfo{author}{\bibfnamefont{M.}~\bibnamefont{Kusakabe}},
  \bibinfo{author}{\bibfnamefont{T.}~\bibnamefont{Kajino}},
  \bibinfo{author}{\bibfnamefont{T.}~\bibnamefont{Yoshida}}, \bibnamefont{and}
  \bibinfo{author}{\bibfnamefont{G.~J.} \bibnamefont{Mathews}},
  \bibinfo{journal}{Phys. Rev.} \textbf{\bibinfo{volume}{D81}},
  \bibinfo{pages}{083521} (\bibinfo{year}{2010}), \eprint{1001.1410}.

\bibitem[{\citenamefont{Cyburt et~al.}(2012)\citenamefont{Cyburt, Ellis,
  Fields, Luo, Olive, and Spanos}}]{Cyburt:2012kp}
\bibinfo{author}{\bibfnamefont{R.~H.} \bibnamefont{Cyburt}},
  \bibinfo{author}{\bibfnamefont{J.}~\bibnamefont{Ellis}},
  \bibinfo{author}{\bibfnamefont{B.~D.} \bibnamefont{Fields}},
  \bibinfo{author}{\bibfnamefont{F.}~\bibnamefont{Luo}},
  \bibinfo{author}{\bibfnamefont{K.~A.} \bibnamefont{Olive}}, \bibnamefont{and}
  \bibinfo{author}{\bibfnamefont{V.~C.} \bibnamefont{Spanos}},
  \bibinfo{journal}{JCAP} \textbf{\bibinfo{volume}{1212}}, \bibinfo{pages}{037}
  (\bibinfo{year}{2012}), \eprint{1209.1347}.

\bibitem[{\citenamefont{Kusakabe et~al.}(2013)\citenamefont{Kusakabe, Kim,
  Cheoun, Kajino, and Kino}}]{Kusakabe:2013tra}
\bibinfo{author}{\bibfnamefont{M.}~\bibnamefont{Kusakabe}},
  \bibinfo{author}{\bibfnamefont{K.~S.} \bibnamefont{Kim}},
  \bibinfo{author}{\bibfnamefont{M.-K.} \bibnamefont{Cheoun}},
  \bibinfo{author}{\bibfnamefont{T.}~\bibnamefont{Kajino}}, \bibnamefont{and}
  \bibinfo{author}{\bibfnamefont{Y.}~\bibnamefont{Kino}},
  \bibinfo{journal}{Phys. Rev.} \textbf{\bibinfo{volume}{D88}},
  \bibinfo{pages}{063514} (\bibinfo{year}{2013}), \bibinfo{note}{[Erratum:
  Phys. Rev.D88,no.8,089904(2013)]}, \eprint{1305.6155}.

\bibitem[{\citenamefont{Kusakabe et~al.}(2014)\citenamefont{Kusakabe, Kim,
  Cheoun, Kajino, Kino, and Mathews}}]{Kusakabe:2014moa}
\bibinfo{author}{\bibfnamefont{M.}~\bibnamefont{Kusakabe}},
  \bibinfo{author}{\bibfnamefont{K.~S.} \bibnamefont{Kim}},
  \bibinfo{author}{\bibfnamefont{M.-K.} \bibnamefont{Cheoun}},
  \bibinfo{author}{\bibfnamefont{T.}~\bibnamefont{Kajino}},
  \bibinfo{author}{\bibfnamefont{Y.}~\bibnamefont{Kino}}, \bibnamefont{and}
  \bibinfo{author}{\bibfnamefont{G.~J.} \bibnamefont{Mathews}},
  \bibinfo{journal}{Astrophys. J. Suppl.} \textbf{\bibinfo{volume}{214}},
  \bibinfo{pages}{5} (\bibinfo{year}{2014}), \eprint{1403.4156}.

\bibitem[{\citenamefont{Yamazaki et~al.}(2014)\citenamefont{Yamazaki, Kusakabe,
  Kajino, Mathews, and Cheoun}}]{Yamazaki:2014fja}
\bibinfo{author}{\bibfnamefont{D.~G.} \bibnamefont{Yamazaki}},
  \bibinfo{author}{\bibfnamefont{M.}~\bibnamefont{Kusakabe}},
  \bibinfo{author}{\bibfnamefont{T.}~\bibnamefont{Kajino}},
  \bibinfo{author}{\bibfnamefont{G.~J.} \bibnamefont{Mathews}},
  \bibnamefont{and} \bibinfo{author}{\bibfnamefont{M.-K.}
  \bibnamefont{Cheoun}}, \bibinfo{journal}{Phys. Rev.}
  \textbf{\bibinfo{volume}{D90}}, \bibinfo{pages}{023001}
  (\bibinfo{year}{2014}), \eprint{1407.0021}.

\bibitem[{\citenamefont{Coc et~al.}(2014)\citenamefont{Coc, Uzan, and
  Vangioni}}]{Coc:2014oia}
\bibinfo{author}{\bibfnamefont{A.}~\bibnamefont{Coc}},
  \bibinfo{author}{\bibfnamefont{J.-P.} \bibnamefont{Uzan}}, \bibnamefont{and}
  \bibinfo{author}{\bibfnamefont{E.}~\bibnamefont{Vangioni}},
  \bibinfo{journal}{JCAP} \textbf{\bibinfo{volume}{1410}}, \bibinfo{pages}{050}
  (\bibinfo{year}{2014}), \eprint{1403.6694}.

\bibitem[{\citenamefont{Mukhamedzhanov
  et~al.}(2016)\citenamefont{Mukhamedzhanov, Shubhchintak, and
  Bertulani}}]{Mukhamedzhanov:2016ecq}
\bibinfo{author}{\bibfnamefont{A.~M.} \bibnamefont{Mukhamedzhanov}},
  \bibinfo{author}{\bibnamefont{Shubhchintak}}, \bibnamefont{and}
  \bibinfo{author}{\bibfnamefont{C.~A.} \bibnamefont{Bertulani}},
  \bibinfo{journal}{Phys. Rev.} \textbf{\bibinfo{volume}{C93}},
  \bibinfo{pages}{045805} (\bibinfo{year}{2016}), \eprint{1602.07395}.

\bibitem[{\citenamefont{Bird et~al.}(2008)\citenamefont{Bird, Koopmans, and
  Pospelov}}]{Bird:2007ge}
\bibinfo{author}{\bibfnamefont{C.}~\bibnamefont{Bird}},
  \bibinfo{author}{\bibfnamefont{K.}~\bibnamefont{Koopmans}}, \bibnamefont{and}
  \bibinfo{author}{\bibfnamefont{M.}~\bibnamefont{Pospelov}},
  \bibinfo{journal}{Phys. Rev.} \textbf{\bibinfo{volume}{D78}},
  \bibinfo{pages}{083010} (\bibinfo{year}{2008}), \eprint{hep-ph/0703096}.

\bibitem[{\citenamefont{Hamaguchi et~al.}(2007)\citenamefont{Hamaguchi,
  Hatsuda, Kamimura, Kino, and Yanagida}}]{Hamaguchi:2007mp}
\bibinfo{author}{\bibfnamefont{K.}~\bibnamefont{Hamaguchi}},
  \bibinfo{author}{\bibfnamefont{T.}~\bibnamefont{Hatsuda}},
  \bibinfo{author}{\bibfnamefont{M.}~\bibnamefont{Kamimura}},
  \bibinfo{author}{\bibfnamefont{Y.}~\bibnamefont{Kino}}, \bibnamefont{and}
  \bibinfo{author}{\bibfnamefont{T.~T.} \bibnamefont{Yanagida}},
  \bibinfo{journal}{Phys. Lett.} \textbf{\bibinfo{volume}{B650}},
  \bibinfo{pages}{268} (\bibinfo{year}{2007}), \eprint{hep-ph/0702274}.

\bibitem[{\citenamefont{Kamimura et~al.}(2009)\citenamefont{Kamimura, Kino, and
  Hiyama}}]{Kamimura:2008fx}
\bibinfo{author}{\bibfnamefont{M.}~\bibnamefont{Kamimura}},
  \bibinfo{author}{\bibfnamefont{Y.}~\bibnamefont{Kino}}, \bibnamefont{and}
  \bibinfo{author}{\bibfnamefont{E.}~\bibnamefont{Hiyama}},
  \bibinfo{journal}{Prog. Theor. Phys.} \textbf{\bibinfo{volume}{121}},
  \bibinfo{pages}{1059} (\bibinfo{year}{2009}), \eprint{0809.4772}.

\bibitem[{\citenamefont{Nardi et~al.}(2006{\natexlab{a}})\citenamefont{Nardi,
  Nir, Racker, and Roulet}}]{Nardi:2005hs}
\bibinfo{author}{\bibfnamefont{E.}~\bibnamefont{Nardi}},
  \bibinfo{author}{\bibfnamefont{Y.}~\bibnamefont{Nir}},
  \bibinfo{author}{\bibfnamefont{J.}~\bibnamefont{Racker}}, \bibnamefont{and}
  \bibinfo{author}{\bibfnamefont{E.}~\bibnamefont{Roulet}},
  \bibinfo{journal}{JHEP} \textbf{\bibinfo{volume}{01}}, \bibinfo{pages}{068}
  (\bibinfo{year}{2006}{\natexlab{a}}), \eprint{hep-ph/0512052}.

\bibitem[{\citenamefont{Nardi et~al.}(2006{\natexlab{b}})\citenamefont{Nardi,
  Nir, Roulet, and Racker}}]{Nardi:2006fx}
\bibinfo{author}{\bibfnamefont{E.}~\bibnamefont{Nardi}},
  \bibinfo{author}{\bibfnamefont{Y.}~\bibnamefont{Nir}},
  \bibinfo{author}{\bibfnamefont{E.}~\bibnamefont{Roulet}}, \bibnamefont{and}
  \bibinfo{author}{\bibfnamefont{J.}~\bibnamefont{Racker}},
  \bibinfo{journal}{JHEP} \textbf{\bibinfo{volume}{01}}, \bibinfo{pages}{164}
  (\bibinfo{year}{2006}{\natexlab{b}}), \eprint{hep-ph/0601084}.

\bibitem[{\citenamefont{Fong et~al.}(2011)\citenamefont{Fong, Gonzalez-Garcia,
  and Nardi}}]{Fong:2011yx}
\bibinfo{author}{\bibfnamefont{C.~S.} \bibnamefont{Fong}},
  \bibinfo{author}{\bibfnamefont{M.~C.} \bibnamefont{Gonzalez-Garcia}},
  \bibnamefont{and} \bibinfo{author}{\bibfnamefont{E.}~\bibnamefont{Nardi}},
  \bibinfo{journal}{Int. J. Mod. Phys.} \textbf{\bibinfo{volume}{A26}},
  \bibinfo{pages}{3491} (\bibinfo{year}{2011}), \eprint{1107.5312}.

\bibitem[{\citenamefont{Ishihara et~al.}(2016)\citenamefont{Ishihara, Maekawa,
  Takegawa, and Yamanaka}}]{Ishihara:2015uua}
\bibinfo{author}{\bibfnamefont{T.}~\bibnamefont{Ishihara}},
  \bibinfo{author}{\bibfnamefont{N.}~\bibnamefont{Maekawa}},
  \bibinfo{author}{\bibfnamefont{M.}~\bibnamefont{Takegawa}}, \bibnamefont{and}
  \bibinfo{author}{\bibfnamefont{M.}~\bibnamefont{Yamanaka}},
  \bibinfo{journal}{JHEP} \textbf{\bibinfo{volume}{02}}, \bibinfo{pages}{108}
  (\bibinfo{year}{2016}), \eprint{1508.06212}.

\bibitem[{\citenamefont{Chung et~al.}(2009)\citenamefont{Chung, Garbrecht,
  Ramsey-Musolf, and Tulin}}]{Chung:2009qs}
\bibinfo{author}{\bibfnamefont{D.~J.~H.} \bibnamefont{Chung}},
  \bibinfo{author}{\bibfnamefont{B.}~\bibnamefont{Garbrecht}},
  \bibinfo{author}{\bibfnamefont{M.}~\bibnamefont{Ramsey-Musolf}},
  \bibnamefont{and} \bibinfo{author}{\bibfnamefont{S.}~\bibnamefont{Tulin}},
  \bibinfo{journal}{JHEP} \textbf{\bibinfo{volume}{12}}, \bibinfo{pages}{067}
  (\bibinfo{year}{2009}), \eprint{0908.2187}.

\bibitem[{\citenamefont{Fong et~al.}(2010)\citenamefont{Fong, Gonzalez-Garcia,
  Nardi, and Racker}}]{Fong:2010qh}
\bibinfo{author}{\bibfnamefont{C.~S.} \bibnamefont{Fong}},
  \bibinfo{author}{\bibfnamefont{M.~C.} \bibnamefont{Gonzalez-Garcia}},
  \bibinfo{author}{\bibfnamefont{E.}~\bibnamefont{Nardi}}, \bibnamefont{and}
  \bibinfo{author}{\bibfnamefont{J.}~\bibnamefont{Racker}},
  \bibinfo{journal}{JCAP} \textbf{\bibinfo{volume}{1012}}, \bibinfo{pages}{013}
  (\bibinfo{year}{2010}), \eprint{1009.0003}.

\bibitem[{\citenamefont{Plumacher}(1998)}]{Plumacher:1997ru}
\bibinfo{author}{\bibfnamefont{M.}~\bibnamefont{Plumacher}},
  \bibinfo{journal}{Nucl. Phys.} \textbf{\bibinfo{volume}{B530}},
  \bibinfo{pages}{207} (\bibinfo{year}{1998}), \eprint{hep-ph/9704231}.

\bibitem[{\citenamefont{Covi et~al.}(1996)\citenamefont{Covi, Roulet, and
  Vissani}}]{Covi:1996wh}
\bibinfo{author}{\bibfnamefont{L.}~\bibnamefont{Covi}},
  \bibinfo{author}{\bibfnamefont{E.}~\bibnamefont{Roulet}}, \bibnamefont{and}
  \bibinfo{author}{\bibfnamefont{F.}~\bibnamefont{Vissani}},
  \bibinfo{journal}{Phys. Lett.} \textbf{\bibinfo{volume}{B384}},
  \bibinfo{pages}{169} (\bibinfo{year}{1996}), \eprint{hep-ph/9605319}.

\bibitem[{\citenamefont{Laine and Shaposhnikov}(2000)}]{Laine:1999wv}
\bibinfo{author}{\bibfnamefont{M.}~\bibnamefont{Laine}} \bibnamefont{and}
  \bibinfo{author}{\bibfnamefont{M.~E.} \bibnamefont{Shaposhnikov}},
  \bibinfo{journal}{Phys. Rev.} \textbf{\bibinfo{volume}{D61}},
  \bibinfo{pages}{117302} (\bibinfo{year}{2000}), \eprint{hep-ph/9911473}.

\bibitem[{\citenamefont{Porod}(2003)}]{Porod:2003um}
\bibinfo{author}{\bibfnamefont{W.}~\bibnamefont{Porod}},
  \bibinfo{journal}{Comput. Phys. Commun.} \textbf{\bibinfo{volume}{153}},
  \bibinfo{pages}{275} (\bibinfo{year}{2003}), \eprint{hep-ph/0301101}.

\bibitem[{\citenamefont{Porod and Staub}(2012)}]{Porod:2011nf}
\bibinfo{author}{\bibfnamefont{W.}~\bibnamefont{Porod}} \bibnamefont{and}
  \bibinfo{author}{\bibfnamefont{F.}~\bibnamefont{Staub}},
  \bibinfo{journal}{Comput. Phys. Commun.} \textbf{\bibinfo{volume}{183}},
  \bibinfo{pages}{2458} (\bibinfo{year}{2012}), \eprint{1104.1573}.

\bibitem[{\citenamefont{Belanger et~al.}(2009)\citenamefont{Belanger, Boudjema,
  Pukhov, and Semenov}}]{Belanger:2008sj}
\bibinfo{author}{\bibfnamefont{G.}~\bibnamefont{Belanger}},
  \bibinfo{author}{\bibfnamefont{F.}~\bibnamefont{Boudjema}},
  \bibinfo{author}{\bibfnamefont{A.}~\bibnamefont{Pukhov}}, \bibnamefont{and}
  \bibinfo{author}{\bibfnamefont{A.}~\bibnamefont{Semenov}},
  \bibinfo{journal}{Comput. Phys. Commun.} \textbf{\bibinfo{volume}{180}},
  \bibinfo{pages}{747} (\bibinfo{year}{2009}), \eprint{0803.2360}.

\bibitem[{\citenamefont{Belanger et~al.}(2011)\citenamefont{Belanger, Boudjema,
  Brun, Pukhov, Rosier-Lees, Salati, and Semenov}}]{Belanger:2010gh}
\bibinfo{author}{\bibfnamefont{G.}~\bibnamefont{Belanger}},
  \bibinfo{author}{\bibfnamefont{F.}~\bibnamefont{Boudjema}},
  \bibinfo{author}{\bibfnamefont{P.}~\bibnamefont{Brun}},
  \bibinfo{author}{\bibfnamefont{A.}~\bibnamefont{Pukhov}},
  \bibinfo{author}{\bibfnamefont{S.}~\bibnamefont{Rosier-Lees}},
  \bibinfo{author}{\bibfnamefont{P.}~\bibnamefont{Salati}}, \bibnamefont{and}
  \bibinfo{author}{\bibfnamefont{A.}~\bibnamefont{Semenov}},
  \bibinfo{journal}{Comput. Phys. Commun.} \textbf{\bibinfo{volume}{182}},
  \bibinfo{pages}{842} (\bibinfo{year}{2011}), \eprint{1004.1092}.

\bibitem[{\citenamefont{Barducci et~al.}(2018)\citenamefont{Barducci, Belanger,
  Bernon, Boudjema, Da~Silva, Kraml, Laa, and Pukhov}}]{Barducci:2016pcb}
\bibinfo{author}{\bibfnamefont{D.}~\bibnamefont{Barducci}},
  \bibinfo{author}{\bibfnamefont{G.}~\bibnamefont{Belanger}},
  \bibinfo{author}{\bibfnamefont{J.}~\bibnamefont{Bernon}},
  \bibinfo{author}{\bibfnamefont{F.}~\bibnamefont{Boudjema}},
  \bibinfo{author}{\bibfnamefont{J.}~\bibnamefont{Da~Silva}},
  \bibinfo{author}{\bibfnamefont{S.}~\bibnamefont{Kraml}},
  \bibinfo{author}{\bibfnamefont{U.}~\bibnamefont{Laa}}, \bibnamefont{and}
  \bibinfo{author}{\bibfnamefont{A.}~\bibnamefont{Pukhov}},
  \bibinfo{journal}{Comput. Phys. Commun.} \textbf{\bibinfo{volume}{222}},
  \bibinfo{pages}{327} (\bibinfo{year}{2018}), \eprint{1606.03834}.

\bibitem[{\citenamefont{Amhis et~al.}(2017)}]{Amhis:2016xyh}
\bibinfo{author}{\bibfnamefont{Y.}~\bibnamefont{Amhis}} \bibnamefont{et~al.}
  (\bibinfo{collaboration}{HFLAV}), \bibinfo{journal}{Eur. Phys. J.}
  \textbf{\bibinfo{volume}{C77}}, \bibinfo{pages}{895} (\bibinfo{year}{2017}),
  \eprint{1612.07233}.

\bibitem[{\citenamefont{Khachatryan et~al.}(2015)}]{CMS:2014xfa}
\bibinfo{author}{\bibfnamefont{V.}~\bibnamefont{Khachatryan}}
  \bibnamefont{et~al.} (\bibinfo{collaboration}{LHCb, CMS}),
  \bibinfo{journal}{Nature} \textbf{\bibinfo{volume}{522}}, \bibinfo{pages}{68}
  (\bibinfo{year}{2015}), \eprint{1411.4413}.

\bibitem[{\citenamefont{Asner et~al.}(2010)}]{Asner:2010qj}
\bibinfo{author}{\bibfnamefont{D.}~\bibnamefont{Asner}} \bibnamefont{et~al.}
  (\bibinfo{collaboration}{Heavy Flavor Averaging Group})
  (\bibinfo{year}{2010}), \eprint{1010.1589}.

\bibitem[{\citenamefont{Bennett et~al.}(2006)}]{Bennett:2006fi}
\bibinfo{author}{\bibfnamefont{G.~W.} \bibnamefont{Bennett}}
  \bibnamefont{et~al.} (\bibinfo{collaboration}{Muon g-2}),
  \bibinfo{journal}{Phys. Rev.} \textbf{\bibinfo{volume}{D73}},
  \bibinfo{pages}{072003} (\bibinfo{year}{2006}), \eprint{hep-ex/0602035}.

\bibitem[{\citenamefont{Davidson and Ibarra}(2002)}]{Davidson:2002qv}
\bibinfo{author}{\bibfnamefont{S.}~\bibnamefont{Davidson}} \bibnamefont{and}
  \bibinfo{author}{\bibfnamefont{A.}~\bibnamefont{Ibarra}},
  \bibinfo{journal}{Phys. Lett.} \textbf{\bibinfo{volume}{B535}},
  \bibinfo{pages}{25} (\bibinfo{year}{2002}), \eprint{hep-ph/0202239}.

\bibitem[{\citenamefont{Akerib et~al.}(2017)}]{Akerib:2016vxi}
\bibinfo{author}{\bibfnamefont{D.~S.} \bibnamefont{Akerib}}
  \bibnamefont{et~al.} (\bibinfo{collaboration}{LUX}), \bibinfo{journal}{Phys.
  Rev. Lett.} \textbf{\bibinfo{volume}{118}}, \bibinfo{pages}{021303}
  (\bibinfo{year}{2017}), \eprint{1608.07648}.

\bibitem[{\citenamefont{Baldini et~al.}(2016)}]{TheMEG:2016wtm}
\bibinfo{author}{\bibfnamefont{A.~M.} \bibnamefont{Baldini}}
  \bibnamefont{et~al.} (\bibinfo{collaboration}{MEG}), \bibinfo{journal}{Eur.
  Phys. J.} \textbf{\bibinfo{volume}{C76}}, \bibinfo{pages}{434}
  (\bibinfo{year}{2016}), \eprint{1605.05081}.

\bibitem[{\citenamefont{Baldini et~al.}(2018)}]{Baldini:2018nnn}
\bibinfo{author}{\bibfnamefont{A.~M.} \bibnamefont{Baldini}}
  \bibnamefont{et~al.} (\bibinfo{collaboration}{MEG II})
  (\bibinfo{year}{2018}), \eprint{1801.04688}.

\bibitem[{\citenamefont{Bellgardt et~al.}(1988)}]{Bellgardt:1987du}
\bibinfo{author}{\bibfnamefont{U.}~\bibnamefont{Bellgardt}}
  \bibnamefont{et~al.} (\bibinfo{collaboration}{SINDRUM}),
  \bibinfo{journal}{Nucl. Phys.} \textbf{\bibinfo{volume}{B299}},
  \bibinfo{pages}{1} (\bibinfo{year}{1988}).

\bibitem[{\citenamefont{Blondel et~al.}(2013)}]{Blondel:2013ia}
\bibinfo{author}{\bibfnamefont{A.}~\bibnamefont{Blondel}} \bibnamefont{et~al.}
  (\bibinfo{year}{2013}), \eprint{1301.6113}.

\bibitem[{\citenamefont{Aubert et~al.}(2010)}]{Aubert:2009ag}
\bibinfo{author}{\bibfnamefont{B.}~\bibnamefont{Aubert}} \bibnamefont{et~al.}
  (\bibinfo{collaboration}{BaBar}), \bibinfo{journal}{Phys. Rev. Lett.}
  \textbf{\bibinfo{volume}{104}}, \bibinfo{pages}{021802}
  (\bibinfo{year}{2010}), \eprint{0908.2381}.

\bibitem[{\citenamefont{Aushev et~al.}(2010)}]{Aushev:2010bq}
\bibinfo{author}{\bibfnamefont{T.}~\bibnamefont{Aushev}} \bibnamefont{et~al.}
  (\bibinfo{year}{2010}), \eprint{1002.5012}.

\bibitem[{\citenamefont{Hayasaka et~al.}(2010)}]{Hayasaka:2010np}
\bibinfo{author}{\bibfnamefont{K.}~\bibnamefont{Hayasaka}}
  \bibnamefont{et~al.}, \bibinfo{journal}{Phys. Lett.}
  \textbf{\bibinfo{volume}{B687}}, \bibinfo{pages}{139} (\bibinfo{year}{2010}),
  \eprint{1001.3221}.

\bibitem[{\citenamefont{Biswas and Mukhopadhyaya}(2009)}]{Biswas:2009zp}
\bibinfo{author}{\bibfnamefont{S.}~\bibnamefont{Biswas}} \bibnamefont{and}
  \bibinfo{author}{\bibfnamefont{B.}~\bibnamefont{Mukhopadhyaya}},
  \bibinfo{journal}{Phys. Rev.} \textbf{\bibinfo{volume}{D79}},
  \bibinfo{pages}{115009} (\bibinfo{year}{2009}), \eprint{0902.4349}.

\bibitem[{\citenamefont{Heisig and Kersten}(2011)}]{Heisig:2011dr}
\bibinfo{author}{\bibfnamefont{J.}~\bibnamefont{Heisig}} \bibnamefont{and}
  \bibinfo{author}{\bibfnamefont{J.}~\bibnamefont{Kersten}},
  \bibinfo{journal}{Phys. Rev.} \textbf{\bibinfo{volume}{D84}},
  \bibinfo{pages}{115009} (\bibinfo{year}{2011}), \eprint{1106.0764}.

\bibitem[{\citenamefont{Hagiwara et~al.}(2013)\citenamefont{Hagiwara, Li,
  Mawatari, and Nakamura}}]{Hagiwara:2012vz}
\bibinfo{author}{\bibfnamefont{K.}~\bibnamefont{Hagiwara}},
  \bibinfo{author}{\bibfnamefont{T.}~\bibnamefont{Li}},
  \bibinfo{author}{\bibfnamefont{K.}~\bibnamefont{Mawatari}}, \bibnamefont{and}
  \bibinfo{author}{\bibfnamefont{J.}~\bibnamefont{Nakamura}},
  \bibinfo{journal}{Eur. Phys. J.} \textbf{\bibinfo{volume}{C73}},
  \bibinfo{pages}{2489} (\bibinfo{year}{2013}), \eprint{1212.6247}.

\bibitem[{\citenamefont{Citron et~al.}(2013)\citenamefont{Citron, Ellis, Luo,
  Marrouche, Olive, and de~Vries}}]{Citron:2012fg}
\bibinfo{author}{\bibfnamefont{M.}~\bibnamefont{Citron}},
  \bibinfo{author}{\bibfnamefont{J.}~\bibnamefont{Ellis}},
  \bibinfo{author}{\bibfnamefont{F.}~\bibnamefont{Luo}},
  \bibinfo{author}{\bibfnamefont{J.}~\bibnamefont{Marrouche}},
  \bibinfo{author}{\bibfnamefont{K.~A.} \bibnamefont{Olive}}, \bibnamefont{and}
  \bibinfo{author}{\bibfnamefont{K.~J.} \bibnamefont{de~Vries}},
  \bibinfo{journal}{Phys. Rev.} \textbf{\bibinfo{volume}{D87}},
  \bibinfo{pages}{036012} (\bibinfo{year}{2013}), \eprint{1212.2886}.

\bibitem[{\citenamefont{Desai et~al.}(2014)\citenamefont{Desai, Ellis, Luo, and
  Marrouche}}]{Desai:2014uha}
\bibinfo{author}{\bibfnamefont{N.}~\bibnamefont{Desai}},
  \bibinfo{author}{\bibfnamefont{J.}~\bibnamefont{Ellis}},
  \bibinfo{author}{\bibfnamefont{F.}~\bibnamefont{Luo}}, \bibnamefont{and}
  \bibinfo{author}{\bibfnamefont{J.}~\bibnamefont{Marrouche}},
  \bibinfo{journal}{Phys. Rev.} \textbf{\bibinfo{volume}{D90}},
  \bibinfo{pages}{055031} (\bibinfo{year}{2014}), \eprint{1404.5061}.

\bibitem[{\citenamefont{Heisig et~al.}(2015)\citenamefont{Heisig, Lessa, and
  Quertenmont}}]{Heisig:2015yla}
\bibinfo{author}{\bibfnamefont{J.}~\bibnamefont{Heisig}},
  \bibinfo{author}{\bibfnamefont{A.}~\bibnamefont{Lessa}}, \bibnamefont{and}
  \bibinfo{author}{\bibfnamefont{L.}~\bibnamefont{Quertenmont}},
  \bibinfo{journal}{JHEP} \textbf{\bibinfo{volume}{12}}, \bibinfo{pages}{087}
  (\bibinfo{year}{2015}), \eprint{1509.00473}.

\bibitem[{\citenamefont{Khoze et~al.}(2017)\citenamefont{Khoze, Plascencia, and
  Sakurai}}]{Khoze:2017ixx}
\bibinfo{author}{\bibfnamefont{V.~V.} \bibnamefont{Khoze}},
  \bibinfo{author}{\bibfnamefont{A.~D.} \bibnamefont{Plascencia}},
  \bibnamefont{and} \bibinfo{author}{\bibfnamefont{K.}~\bibnamefont{Sakurai}},
  \bibinfo{journal}{JHEP} \textbf{\bibinfo{volume}{06}}, \bibinfo{pages}{041}
  (\bibinfo{year}{2017}), \eprint{1702.00750}.

\end{thebibliography}

\end{document}